\documentclass[%
 aip,
 amsmath,amssymb,
 reprint,%
]{revtex4-1}

\usepackage{graphicx}% Include figure files
\usepackage{dcolumn}% Align table columns on decimal point
\usepackage{bm}% bold math
\usepackage[utf8]{inputenc}
\usepackage[T1]{fontenc}
\usepackage{mathptmx}
\usepackage{etoolbox}
\usepackage{caption}
\usepackage{subcaption}
\usepackage{natbib}
\usepackage{xcolor}

\usepackage{xr}

\makeatletter
\newcommand*{\addFileDependency}[1]{% argument=file name and extension
\typeout{(#1)}% latexmk will find this if $recorder=0
\@addtofilelist{#1}
\IfFileExists{#1}{}{\typeout{No file #1.}}
}\makeatother

\newcommand*\stref[1]{%
    Supplementary Table \ref{#1}}
 
\newcommand*\ssref[1]{%
    Supplementary Section S\ref{#1}}

\makeatletter
\def\@email#1#2{%
 \endgroup
 \patchcmd{\titleblock@produce}
  {\frontmatter@RRAPformat}
  {\frontmatter@RRAPformat{\produce@RRAP{*#1\href{mailto:#2}{#2}}}\frontmatter@RRAPformat}
  {}{}
}%
\makeatother
\begin{document}

%\preprint{AIP/123-QED}

\title[Janus helices]{Janus helices: From fully attractive to hard helices}
% Force line breaks with \\

\author{Laura Dal Compare}
\email{laura.dalcompare@unive.it}
\affiliation{Dipartimento di Scienze Molecolari e Nanosistemi, 
Universit\`{a} Ca' Foscari di Venezia
Campus Scientifico, Edificio Alfa,
via Torino 155,30170 Venezia Mestre, Italy}

\author{Flavio Romano}
\email{flavio.romano@unive.it}
\affiliation{Dipartimento di Scienze Molecolari e Nanosistemi, 
Universit\`{a} Ca' Foscari di Venezia
Campus Scientifico, Edificio Alfa,
via Torino 155,30170 Venezia Mestre, Italy}
\affiliation{European Centre for Living Technology (ECLT)
Ca' Bottacin, 3911 Dorsoduro Calle Crosera, 
30123 Venice, Italy}

% \altaffiliation[Also at ]{European Centre for Living Technology (ECLT)
%Ca' Bottacin, 3911 Dorsoduro Calle Crosera, 
%30123 Venice, Italy}%Lines break automatically or can be forced with \\
\author{Jared A. Wood}%
 \email{jared.wood@sydney.edu.au}
\affiliation{ ARC Centre of Excellence in Exciton Science, School of Chemistry, University of Sydney, Sydney, New South Wales 2006, Australia%\\This line break forced with \textbackslash\textbackslash
}%
\affiliation{The University of Sydney Nano Institute, University of Sydney, New South Wales 2006, Australia}

\author{Asaph Widmer-Cooper}
 %\homepage{http://www.Second.institution.edu/~Charlie.Author.}
 \email{asaph.widmer-cooper@sydney.edu.au}
 \affiliation{ ARC Centre of Excellence in Exciton Science, School of Chemistry, University of Sydney, Sydney, New South Wales 2006, Australia%\\This line break forced with \textbackslash\textbackslash
}%
\affiliation{The University of Sydney Nano Institute, University of Sydney, New South Wales 2006, Australia}

\author{Achille Giacometti}
\email{achille.giacometti@unive.it}
\affiliation{Dipartimento di Scienze Molecolari e Nanosistemi, 
Universit\`{a} Ca' Foscari di Venezia
Campus Scientifico, Edificio Alfa,
via Torino 155,30170 Venezia Mestre, Italy}
\affiliation{European Centre for Living Technology (ECLT)
Ca' Bottacin, 3911 Dorsoduro Calle Crosera, 
30123 Venice, Italy}
\date{\today}% It is always \today, today,

\begin{abstract}
The phase diagram of hard helices differs from its hard rods counterpart by the presence of chiral "screw" phases stemming from the characteristic helical shape, in addition to the conventional liquid crystal phases also found for rod-like particles. Using extensive Monte Carlo and Molecular Dynamics simulations, we study the effect of the addition of a short-range attractive tail representing solvent-induced interactions to a fraction of the sites forming the hard helices, ranging from a single-site attraction to fully attractive helices for a specific helical shape. Different temperature regimes exist for different fractions of the attractive sites, as assessed in terms of the relative Boyle temperatures, that are found to be rather insensitive to the specific shape of the helical particle. The temperature range probed by the present study is well above the corresponding Boyle temperatures, with the phase behaviour still mainly entropically dominated and with the existence and location of the various liquid crystal phases only marginally affected. The pressure in the equation of state is found to decrease upon increasing the fraction of attractive beads and/or on lowering the temperature at fixed volume fraction, as expected on physical grounds.
All screw phases are found to be stable within the considered range of temperatures with the smectic phase becoming more stable on lowering the temperature. By contrast, the location of the transition lines do not display a simple dependence on the fraction of attractive beads in the considered range of temperatures.
\end{abstract}

\maketitle

%\begin{quotation}
%The ``lead paragraph'' is encapsulated with the \LaTeX\ 
%\verb+quotation+ environment and is formatted as a single paragraph before the first section %heading. 
%(The \verb+quotation+ environment reverts to its usual meaning after the first sectioning %command.) 
%Note that numbered references are allowed in the lead paragraph.
%%
%The lead paragraph will only be found in an article being prepared for the journal %\textit{Chaos}.
%\end{quotation}

%%%%%%%%%%%%%%%%%%%%%%%%%%%%%%%%%%%%%%%%%%%%%%%%%%%%%%%%%%%%%%%%%%%
\section{Introduction}
\label{sec:introduction}
%%%%%%%%%%%%%%%%%%%%%%%%%%%%%%%%%%%%%%%%%%%%%%%%%%%%%%%%%%%%%%%%%%%
Guided by site-specific directional interactions, colloidal patchy particles have been shown, both numerically \cite{Glotzer2007,Damasceno2012} and experimentally \cite{Walther2013}, to self-assemble into higher order structures that cannot be obtained by other synthetic routes, thus paving the way to obtain unique morphologies for various technological and biomedical applications. Among the key factors controlling the phases and the morphologies of the self-assembled structure are the location, number, and size of the patches, as well as the strength and range of interactions. Together with thermodynamic conditions, such as the effective temperature and pressure, the number of patches and their arrangement form a relatively small set of parameters that can be tuned to guide the self-assembly toward the desired structure.

Janus colloids \cite{Sciortino2009,Sciortino2010,Chen2011b} are spherical particles having one attractive patch on an otherwise repulsive particle surface that demonstrate how the heterogeneity and directionality of interactions can be designed to achieve a prescribed bulk material with specific properties. Notwithstanding the challenges in the synthesis and functionalization of such colloids, several techniques have been devised that demonstrate the ability to control these processes. A spectacular example of this is provided by the experimental self-assembly of spherical triblock Janus particles into a two-dimensional kagome lattice \cite{Chen2011a}, matched by numerical analysis \cite{Romano2011}.

Within the general class of Janus colloids where the coverage, \textit{i.e.}, the fraction of attractive surface patch compared to the total surface, can be tuned from $0\%$ (hard spheres) to $100\%$ (fully-attractive hard spheres), the Janus limit of equal coverage with just two patches of equal size, one attractive and one repulsive ($50\%$ coverage), has been shown to display a particularly interesting behaviour \cite{Sciortino2009,Sciortino2010,Giacometti2014}. For this system, a gas of micelles is formed at low temperature and low density that has lower energy than the coexisting homogeneous liquid formed at the same temperature and higher density. The corresponding gas-liquid coexistence curve in the pressure-temperature plane is negatively sloped, mimicking the anomalous behavior found in water~\cite{Gallo2016}.

While controlling the interactions plays a paramount role in guiding the self-assembly process, the role of excluded-volume interactions provided by the shape of the colloidal particles has been recognized to be equally important, especially at high density \cite{Damasceno2012}. Surprisingly complex and remarkably diverse colloidal crystals can self-assemble driven purely by particle shape anisotropy and the tendency to maximize the entropy, with this somewhat counter-intuitive mechanism effectively behaving as an entropic bond \cite{Harper2019}. The combination of directional interactions and shape anisotropy is clearly possible, and examples in the literature include colloidal dumbbells \cite{Avvisati2015}, "Mickey Mouse" shaped colloids \cite{Wolters2015}, Janus dumbbells \cite{Munao2015,OToole2017a,OToole2017b}, rods \cite{Chaudhary2012}, ellipsoids \cite{Liu2012,Xu2015} and lobed particles \cite{Paul2020a,Paul2020b}, and polyhedral nanoparticles \cite{Henzie2012}.

Colloidal helices are another example of colloids that can be synthesized \cite{Oh2019} and designed from first principles \cite{Fejer2014}. Simple hard helices formed by spherical fused beads (colloids) arranged into a helical shape with tunable pitch and radius represents a paradigmatic example of entropically driven self-assembly providing a rich and unconventional behaviour \cite{Frezza2013,Kolli2014a,Kolli2014b,Frezza2014,Dussi2015,Kolli2016,Cinacchi2017} that can also find biological representation in colloidal suspensions of semiflexible virus particles \cite{Dogic1997,Dogic2006,Grelet2014}. Elongated hard helices with large aspect ratio form liquid crystal phases that are similar to their hard rods counterpart. Unlike hard rods, however, hard helices are chiral objects and hence can form chiral phases (e.g. a chiral nematic also known as a cholesteric phase) whose origin can be traced back to the shape and the chirality of the helices. More specifically, it was found that the nematic phase was either cholesteric, where the main director is rotating about an axis perpendicular to the original plane, or screw-nematic, where the secondary director is rotating about an axis parallel to the main director \cite{Frezza2014,Dussi2015,Cinacchi2017}. Even the smectic phase was found to display screw-like behaviour under some circumstances \cite{Kolli2014a,Kolli2014b,Kolli2016}. The entropy-driven formation of chiral nematic phases is not limited to helical shapes but includes other interesting examples, such as particles with twisted polyhedral shape \cite{Dussi2016} and curled hard-rods \cite{Wensink2015}.

The aim of the present work is to study Janus helices, which combine enthalpic bonding guided by directional interactions with the tendency of helices to maximize the number of available microstates driven by entropy. This is a natural extension of previous work on the phase behavior of Janus spheres \cite{Sciortino2009,Sciortino2010} and hard helices \cite{Frezza2013,Kolli2014a,Kolli2014b,Kolli2016,Cinacchi2017}. As in the case of Janus spheres, the attractive sites represents interactions induced by the presence of the solvent on different moieties of the helix.  Starting from the case of hard helices, we progressively add a short-range attraction, in the form of a square-well tail, to some of the beads forming the helices up to the case where all beads are attractive.  The main aim of the present study is to discuss the effect on the liquid-crystal phases displayed by hard helices with the addition of a weak and short-range attraction to some of the beads forming the helices; and at the same time to extend previous Monte Carlo simulations based on hard core interactions with the addition of Molecular Dynamics simulations that allow for faster equilibration and more extensive analysis at high volume fraction. The Janus helices considered in the present study have not been studied previously, but some comparison can be made with the behaviour of fully attractive \cite{Gamez2017} and Janus rods \cite{Tripathy2013} in the limit of very slender helices. The case of a weak and highly localized directional attractive site is also of experimental interest and has already been studied both experimentally and numerically \cite{Chaudhary2012,Repula2019,Jack2021}. 

To set the stage, it is important to identify the regimes to be probed. For spherical Janus colloids, the interesting regime occurs at very low temperatures and low densities where self-assembly competes with the possibility of observing gas-liquid and liquid-solid transitions. For athermal hard helices, the interesting regime occurs at high pressures (and hence high densities) where the isotropic phase is progressively replaced by liquid crystal phases, having positional disorder but orientational order.
Before probing the challenging combined regime of low temperature and high pressure, it is desirable to understand the effect of including a temperature dependence (through the addition of an attractive part to the potential) on the hard helices liquid crystal phases, as well as the effect of the coverage. This will be systematically done in the present study by considering coverages $\chi$ ranging from hard helices ($0\%$ $\chi=0$) to fully attractive square-well (SW) helices ($\chi=100\%$), and includes the single site attractive case ($\chi=6.7\%$ coverage), and the Janus limit ($\chi=50\%$). Particular emphasis will be devoted to understanding the stability of the peculiar screw-like nematic phase that, so far, has only been found for hard helices. The complementary limit of low temperature and low density (or pressure) will be discussed in a companion paper \cite{Wood2023} that will focus on Janus rods, thus removing the additional complexity stemming from the chirality of the helices. 
The shape dependence is also known to play an important role in the case of hard helices \cite{Kolli2014a,Kolli2014b,Dussi2015,Cinacchi2017}. Hence, the present study will be confined to the case of slender helices that can be more easily contrasted with results from the Janus rods.  

The outline of the paper is as follows. Section \ref{sec:theory} presents the model (Section \ref{subsec:Model}) and the required order parameters and correlation functions (Section \ref{subsec:order}). In order to identify the temperature regime far from a possible gas-liquid transition and clearly study competition between self-assembly and liquid crystal formation, a preliminary calculation of the second virial and the corresponding Boyle temperature (the temperature of vanishing second virial coefficient) is carried out in Section \ref{subsec:B2_chi} as a function of the coverage. Most of the successive analyses presented in Section \ref{sec:results} will then refer to temperatures well above the Boyle temperatures, where entropic effects due to excluded volume dominate, with a special focus on the location of the liquid crystal phases and the effect of the temperature and increasing attraction on the equation of state.
Finally, in Section \ref{sec:conclusions} we summarise the key findings from this work and discuss directions for future work.
%%%%%%%%%%%%%%%%%%%%%%%%%%%%%%%%%%%%%%%%%%%%%%%%%%%%%%%%%%%%%%%%%%%%
\section{Theory and Methods}
\label{sec:theory}
%%%%%%%%%%%%%%%%%%%%%%%%%%%%%%%%%%%%%%%%%%%%%%%%%%%%%%%%%%%%%%%%%%%%%%
%%%%%%%%%%%%%%%%%%%%%%%%%%%%%%%%%%%%%%%%%%%%%%%%%%%%%%%%%%%%%%%%%%%%%%
\subsection{Model}
\label{subsec:Model}
%%%%%%%%%%%%%%%%%%%%%%%%%%%%%%%%%%%%%%%%%%%%%%%%%%%%%%%%%%%%%%%%%%%%%%%
%%%%%%%%%%%%%%%%%%%%%%%% Fig 1 %%%%%%%%%%%%%%%%%%%%%%%%%%%%%%%
\begin{figure}[htpb]
\centering
\captionsetup{justification=raggedright,width=1.0\linewidth}
\includegraphics[trim=0 90 0 0,clip,width=0.6\linewidth]{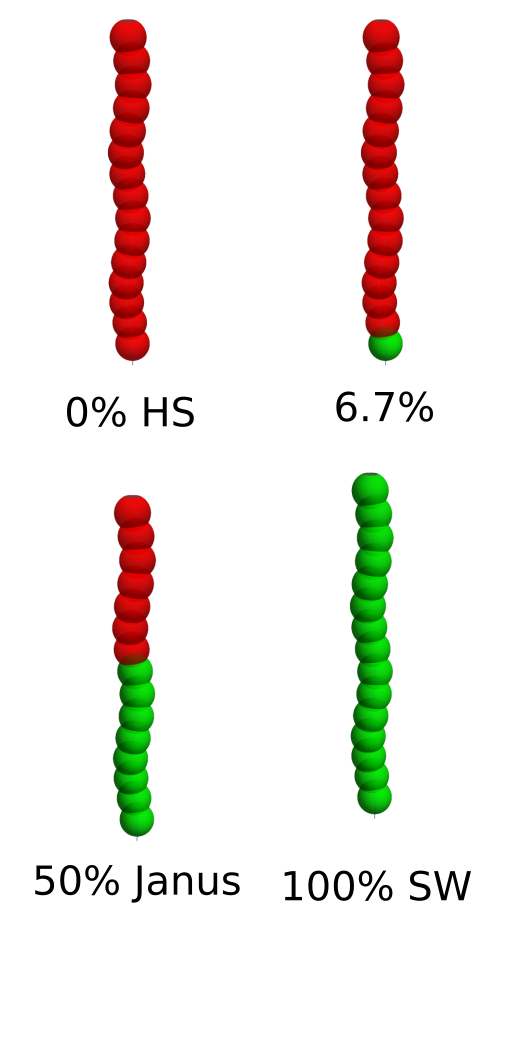}% Here is how to import EPS art
\caption{From hard helices to fully attractive square-well helices. Red beads have HS potential, green beads SW potential. The helices shown have $R=0.2D$ and $p=8D$ with a train of $n_s=15$ spherical beads arranged in a helix with contour length $L=10D$, where $D$ is the diameter of each bead, which is the case considered in the present study. Intermediate cases correspond to 1/15 ($\approx 6.7\%$) and 8/15 ($\approx 50\%$) attractive sites. The $50\%$ case corresponds to the Janus limit.}
\label{fig:fig1} 
\end{figure}
%%%%%%%%%%%%%%%%%%%%%%%%%%%%%%%%%%%%%%%%%%%%%%%%%%%%%%%%%%%%%%%
Hard helices have been implemented as 
a line of 15 fused hard spherical beads of diameter $D$ (the unit of length)
rigidly arranged into a helicoidal shape, with a contour length fixed to $L$. 
On changing the radius $R$ and pitch $p$ at fixed total contour length $L$, 
the shape of the helical particle can be tuned  
from a straight rod to very wound coils. 

We define the coverage $0\le \chi\le 1$ as the fraction of attractive sites, with $\chi=0$ ($0\%$ coverage) representing the HS case, $\chi=0.5$ ($50\%$ coverage) representing the Janus limit, and $\chi=1.0$ ($100\%$ coverage) the SW fully attractive case. Figure \ref{fig:fig1} shows the progressive increase of the coverage in the specific case of a slender helix with $R=0.2$ $p=8D$.

Following past studies \cite{Kolli2014a,Kolli2014b,Kolli2016}, we can organize the shapes into three different classes - small, medium, and large curliness - identified by the
radius $R$ and pitch $p$ of the helical shape, that are related at a fixed total contour length \cite{Frezza2013}. In the present study, we focus on the case of small curliness corresponding to $R=0.2 D$ and $p=8D$, as it can be easily related to its Janus rod counterpart.
The contour length has been fixed to $L=10D$, which is sufficiently large to probe all of the liquid crystal phases for both hard spherocylinders \cite{Bolhuis1997} and hard cylinders \cite{Lopes2021}. As in our previous studies \cite{Kolli2014a,Kolli2014b,Kolli2016}, we have assumed $n_s=15$ spherical beads forming the helices, some of which will be made attractive as detailed below.

Two complementary methods have been used to study the phase behaviour of this system. The first is isobaric-isothermal (NPT) Monte Carlo (MC) simulations using a shape-adapting rectangular box with periodic boundary conditions, which is an extension of the approach used in previous investigations of hard helices \cite{Frezza2013,Kolli2014a,Kolli2014b,Kolli2016}.
Here, the excluded volume between different beads belonging to different helices separated by a distance $r$ is modelled as a pure hard-sphere (HS) potential
\begin{equation}
\phi_{\text{HS}}\left(r\right)=\begin{cases}
+\infty\,,\quad \,r < D &\\
0, \qquad r \geq D&
\end{cases}
\label{sec1:eq1}
\end{equation}
and this was the only interaction appearing in past work.
When attraction is added, we assume that two attractive beads belonging to different helices will interact via a square well (SW) interaction 
\begin{equation}
\phi_{\text{SW}}\left(r\right)=\begin{cases}
+\infty\,,\quad \,r < D &\\
-\epsilon/n_s^2,\quad\, D \leq r < R_c&\\
0, \qquad r \geq R_c&
\end{cases}
\label{sec1:eq2}
\end{equation}
where an attraction having strength $-\epsilon/n_s^2$ is added in the presence of a favourable bond and no corresponding overlap. The range of attraction $R_c$ clearly plays a fundamental role, especially for non-convex objects such as helices: when $R_c$ is large, all attractive beads on different helices are able to interact, whereas only single pairs of beads interact in the presence of very short-range attraction. As in the case of Janus spheres \cite{Sciortino2009,Sciortino2010}, we expect an intermediate regime given by $R_c=1.5D$ to be the most representative, and this value will be used in all NPT-MC simulations of the present study. 

It is worth stressing that each interaction is normalized by the maximum possible number of pair-wise interacting spheres $n_s^2$ so that for a $100\%$ (SW) case, $\epsilon$ is the  energy scale of the problem.
%This choice is consistent with the Janus sphere counterpart where upon decreasing the fraction of attractive sites (i.e. the coverage), the total attraction energy decreases from $\epsilon$ before vanishing for $0\%$ (HS) attraction. 
This choice allows to more easily compare helices with different number of sites.

%In these NPT-MC simulations, with constant temperature $T$ and pressure $P$, each MC move consists of a cycle where each of the $N$ helices is tested, on average, for a possible move, and a single volume change at move where the volume of the computational box is attempted to change at a given pressure.
In NPT-MC simulations, at constant temperature $T$ and constant pressure $P$, each MC move consists of $N$ attempts to move a randomly-selected particle and a single volume change attempt. A particle move attempt consists of simultaneous random translation of the center of mass and a random rotation about it. We have used two sets of simulations with different numbers of helices to test for finite size effects. In the "small-size" simulations, we used $N=972$ helices with $3 \times 10^6$ steps for equilibration and an additional $10^6$ steps as a production run, in line with previous simulations of hard helices \cite{Frezza2013,Kolli2014a,Kolli2016}. In the "large-size" simulations, we used $N=2400$ helices with an equilibration time of $6\times 10^6$ MC steps and an additional $10^6$ steps to collect statistics. The initial condition was a low-density array of parallel helices, with all the attractive parts (if any) initially pointing along the $+z$ "up" directions. 
We also performed Molecular Dynamics (MD) simulations at constant volume and temperature (NVT) for $N=4068$ helices using the software package LAMMPS \cite{PlimptonJComputPhys1995} employing the Nos\'e-Hoover thermostat \cite{HooverPRA1985} to keep the temperature fixed.
For MD simulations, we used a combination of a Weeks-Chandler-Anderson (WCA) potential~\cite{Weeks1971} to mimic hard-core repulsion and a pseudo-square-well (PSW) potential~\cite{ZeronMolecPhys2018} to mimic attraction:

\begin{equation}
\phi(r) = \begin{cases}
\phi_{\text{PSW}}(r) \quad \text{between attractive sites} \\
\phi_{\text{WCA}}(r) \quad \text{otherwise} 
\end{cases}
\end{equation}
where
%\begin{eqnarray}
%\label{sec1:eq3a}
%\phi\left(r\right) &=& \phi_{WCA}\left(r\right)+\phi_{\text{PSW}} \left(r\right) 
%\end{eqnarray}

\begin{eqnarray}
\label{sec1:eq3}
\phi_{WCA}\left(r\right) &=& 4 \epsilon \left[\left(\frac{D}{r}\right)^{12}- \left(\frac{D}{r}\right)^{6} + \frac{1}{4} \right]
\end{eqnarray}
which vanishes for $r\ge \sqrt[6]{2} D$ and acts on the repulsive-repulsive and repulsive-attractive bead pairs, and

\begin{eqnarray}
\label{sec1:eq4}
\phi_{\text{PSW}} \left(r\right) &=& \frac{\epsilon}{2}\left[\left(\frac{D}{r}\right)^{n}
+ \frac{1-e^{-\lambda\left(r/D-1\right)\left((r-R_c)/D\right)}}{1+e^{-\lambda\left(r/D-1\right)\left((r-R_c)/D\right)}} -1 \right] 
\end{eqnarray}
which act only among the attractive beads, where $n$ and $\lambda$ are parameters that tune the shape of the well. Here we have used $\lambda=50$ and $n=200$ so that $\phi_{PSW}(r)$ approximates a square-well tail while allowing for efficient equilibration.
This is the same potential used in the companion paper \cite{Wood2023}, where it is shown to reproduce the phase behavior of the SW-line model \cite{Gamez2017} to within a small ($<5\%$) shift of the phase boundaries to higher volume fraction.
Slightly different potentials have been used to model similar systems \cite{DeBraaf2017,LiuJCP2019,Liu2022}, but we expect the results to depend only weakly on the details of the model potential. Natural reduced units are $T^{*}=k_BT/\epsilon$ for temperature and for pressures $P D^3/k_BT$ in the case of HS interactions and $P D^3/\epsilon$ in the case with attractions. We will also use the fraction of the total volume occupied by the helices, $\eta=V_{occ}/V$, as a measure of their concentration. Here $V_{occ}=v_0 N$, where $v_0$ is the volume of a single helix and $N$ is the number of helices \cite{Frezza2013}. Our time unit in the NVT-MD analysis is $(m/k_B T)^{1/2} D$, where $m$ is the mass of each bead that is taken so that the mass of the helix has a unit value. As in the MC case, the initial configuration was a crystal phase with all helices perfectly aligned in an AAA stack with all of the attractive patches directed along the positive $z$ direction.
\textcolor{black}{To test for possible dependence on the initial conditions, we also considered a situation where the attractive beads were aligned up in one layer and down in the immediately successive layer, so that the attractive beads on the helices belonging to the two layers were initially in contact.}

Snapshots are color-coded according to the helix orientation in the case of hard (HS) helices (as in previous work) \cite{Kolli2014a,Kolli2014b,Kolli2016,Cinacchi2017}, as well as for fully attractive (SW) helices, and according to the coverage (green attractive, red repulsive sites) in the case of partially attractive helices, in analogy with the spherical counterpart  \cite{Sciortino2009,Sciortino2010}.
%%%%%%%%%%%%%%%%%%%%%%%%%%%%%%%%%%%%%%%%
\subsection{Order parameters and screw-like liquid crystal phases}
\label{subsec:order}
%%%%%%%%%%%%%%%%%%%%%%%%%%%%%%%%%%%%%

%%%%%%%%%%%%%%%%%%%%%%%% Fig 2 %%%%%%%%%%%%%%%%%%%%%%%%%%%%%%%
\begin{figure}[htpb]
  \centering
   \begin{subfigure}{4.0cm}
  \captionsetup{justification=raggedright,width=\linewidth}
    \includegraphics[trim=0 120 0 0,clip,width=1.0\linewidth]{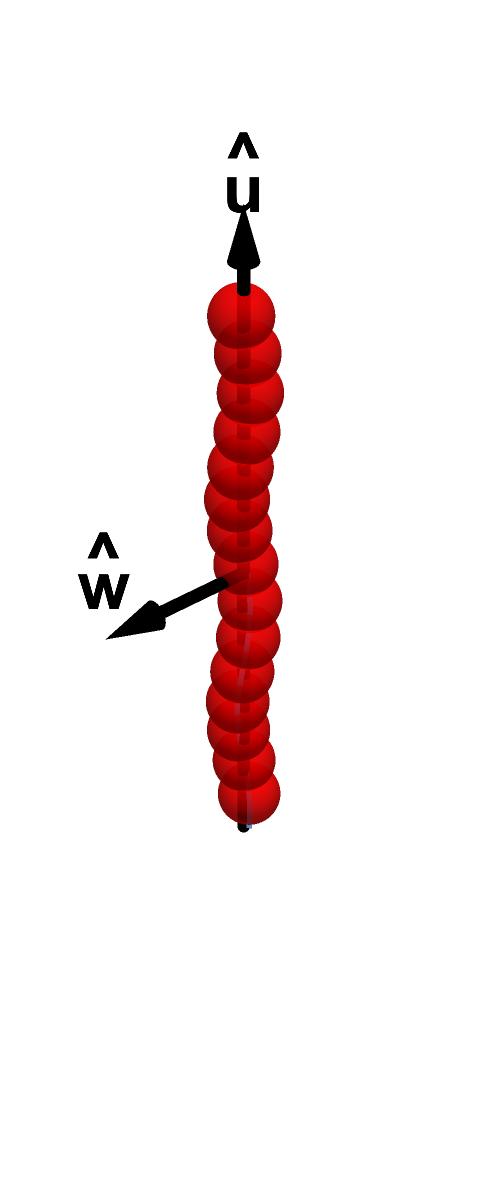}
     \caption{}\label{fig:fig2a}
   \end{subfigure}
  \begin{subfigure}{4.0cm}
  \captionsetup{justification=raggedright,width=\linewidth}
    \includegraphics[trim=0 15 0 0,clip,width=1.0\linewidth]{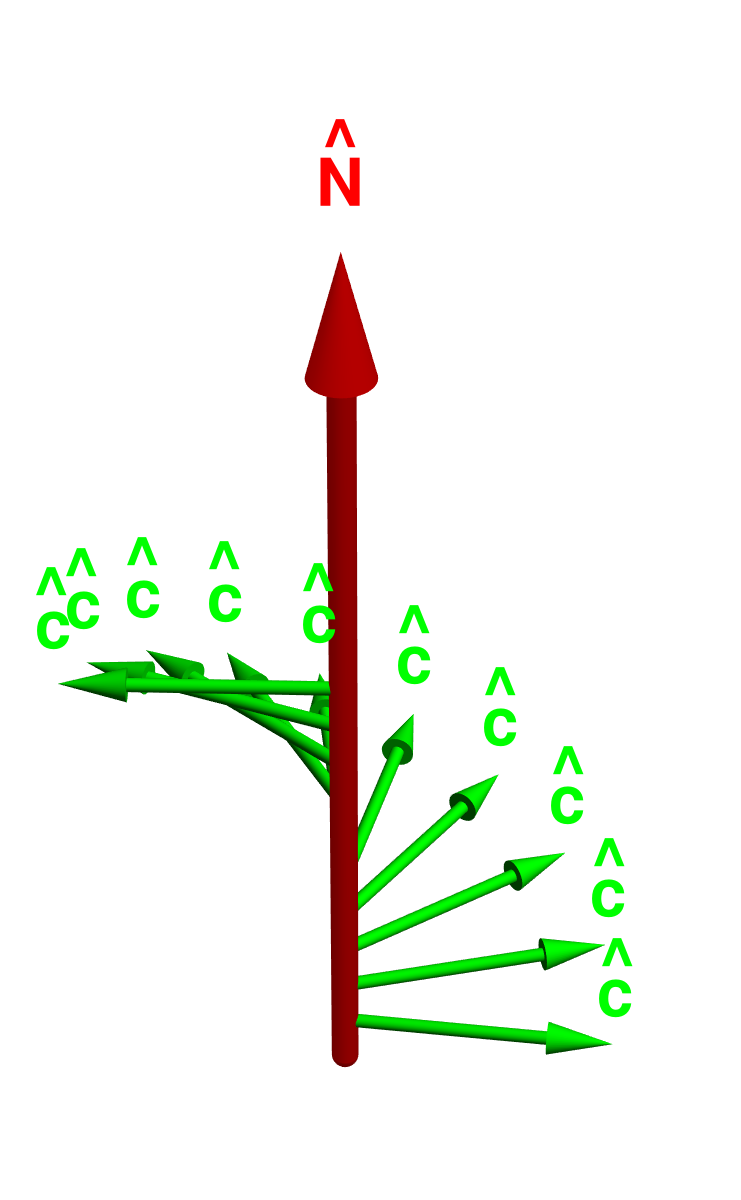}
     \caption{}\label{fig:fig2b}
   \end{subfigure}
   \begin{subfigure}{4.5cm}
    \includegraphics[trim=10 50 10 30,clip,width=1.0\linewidth]{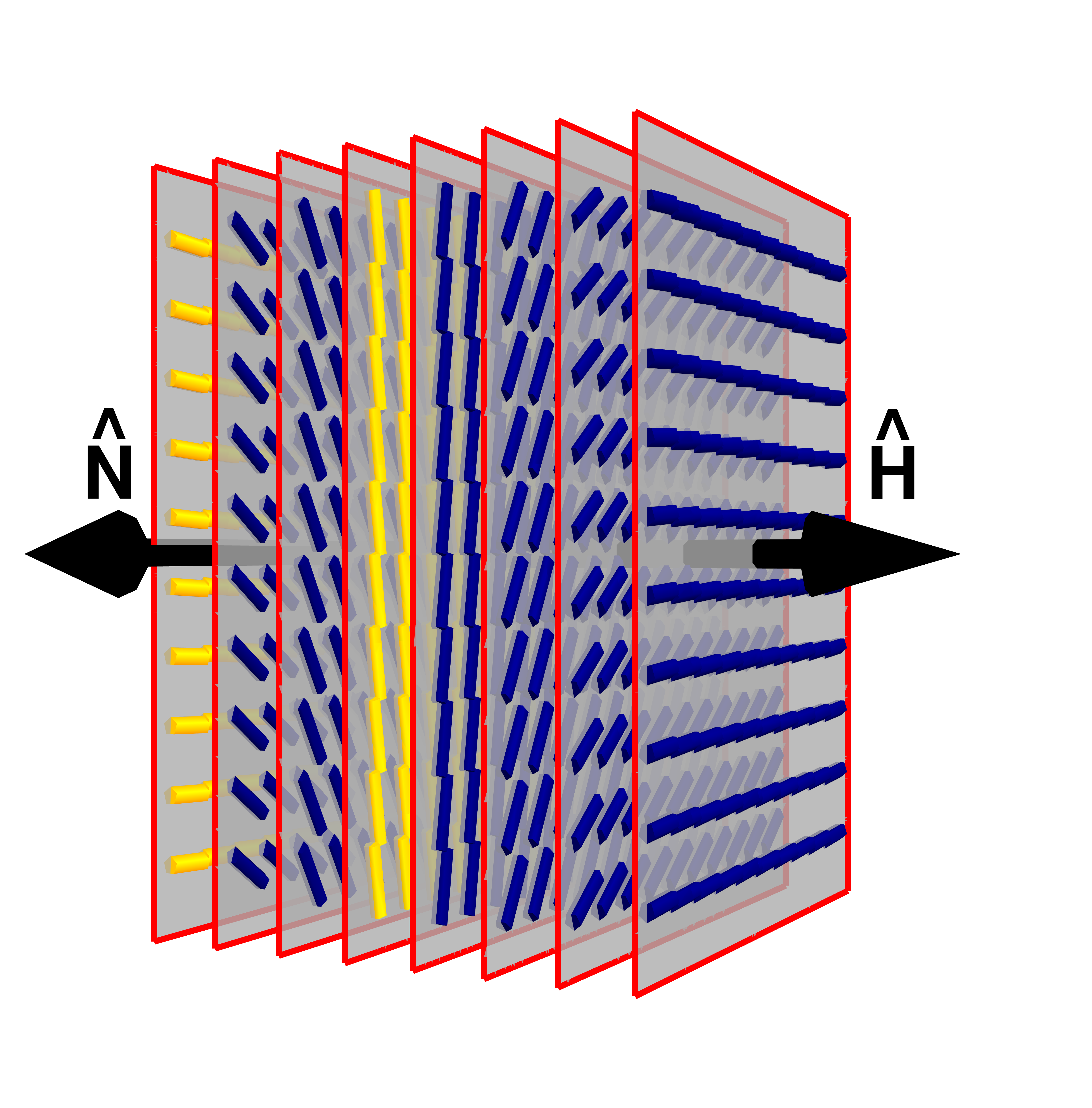}
    \caption{}\label{fig:fig2c}
   \end{subfigure}
   \begin{subfigure}{4.0cm}
    \includegraphics[trim=0 0 0 0,clip,width=1.0\linewidth]{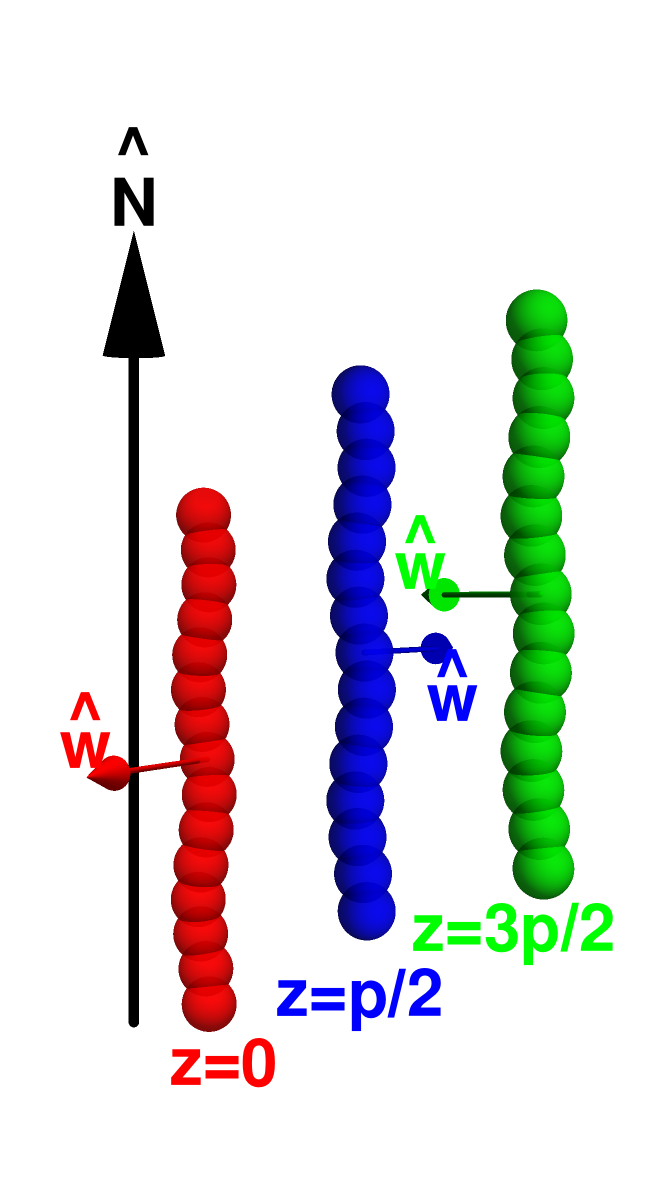}
     \caption{}\label{fig:fig2d}
   \end{subfigure}
  \caption{(a) The local frame for the helical model used in this work \cite{Frezza2013}. Here $\widehat{\mathbf{u}}$ and $\widehat{\mathbf{w}}$ identify the main and secondary helical axes; (b)  The screw-like phase where the secondary director $\widehat{\mathbf{C}}$ rotates in a helical fashion about the main director $\widehat{\mathbf{N}}$; (c) The cholesteric phase where the main director $\widehat{\mathbf{N}}$ rotates in a helical fashion about an axis  $\widehat{\mathbf{H}}$ that is perpendicular to the original plane; (d) The coupling between translation along the main director $\widehat{\mathbf{N}}$ and rotation of the secondary director $\widehat{\mathbf{C}}$ characteristic of the screw-like phase. The three helices have a relative vertical displacement of $z=0$, $z=p/2$, and $z=3p/2$ respectively, where $p$ is the pitch of the helix and hence their $\widehat{\mathbf{w}}$ directors rotate by $\pi/4$ about the main director $\widehat{\mathbf{N}}$ every $p/2$. Note that in this case their main axes $\widehat{\mathbf{u}}$ are all parallel to the phase main director $\widehat{\mathbf{N}}$. All cases shown here refer to the most slender helix considered with $R=0.2D$ and $p=8D$.}
  \label{fig:fig2}
\end{figure}
%%%%%%%%%%%%%%%%%%%%%%%%%%%%%%%%%%%%%%%%%%%%%%%%%%%%%%%%%%%%%%%
The presence of different phases with specific features requires the definition of order parameters able to distinguish between them. To this aim, we have used a combination of global order parameters and specific correlation functions that can act as local order parameters \cite{Lopes2021} that have been introduced recently in the literature. In the following, we will be using capital letters ($\widehat{\mathbf{N}}$ and $\widehat{\mathbf{C}}$), to identify principal directions of the liquid crystal phases, and lower case letters for local axes identifying the orientation of the helices.
The position and orientation of each helix in space is identified by its center of mass $\mathbf{r}$ and by three unit vectors $\widehat{\mathbf{u}}$, $\widehat{\mathbf{v}}$, and $\widehat{\mathbf{w}}$. Here, $\widehat{\mathbf{u}}$ is identifying the helix main axis, whereas  $\widehat{\mathbf{w}}$ and $\widehat{\mathbf{v}}$ are two unit vectors, perpendicular to each other and to $\widehat{\mathbf{u}}$, that identify the azimuthal orientation of the helix in this perpendicular plane. As only one of the two is required for this aim, we will focus on $\widehat{\mathbf{w}}$ in the following (see Figure \ref{fig:fig2a}).
The onset of a nematic phase is best identified by evaluating the Veilliard-Baron tensor \cite{Vieillard1974} 
\begin{eqnarray}
  \label{sec1:eq3a}
  \mathbf{Q}_{\widehat{\mathbf{u}}} &=& \left \langle \frac{1}{N} \sum_{j=1}^{N} \left[\frac{3}{2} \widehat{\mathbf{u}}_{j} \widehat{\mathbf{u}}_{j} -\frac{1}{2} \mathbf{I} \right]  \right \rangle
\end{eqnarray}
where $\widehat{\mathbf{u}}_i$ is the unit vector identifying the orientation of the $i-$th helix in space.
The maximum eigenvalue $\Lambda_{u_1}$ gives the nematic order parameter $\langle P_2 \rangle$, $\approx 1$ for a nematic phase and $\approx 0$ for an isotropic phase, and the corresponding eigenvector gives the nematic director $\widehat{\mathbf{N}}$ (see Figure \ref{fig:fig2b}). Because of the uniaxial symmetry of the helix and the traceless character of tensor $\mathbf{Q}_{\widehat{\mathbf{u}}}$, the other two eigenvalues $\Lambda_{u_2}=\Lambda_{u_3}$ will be identical and negative. For hard helices, an additional chiral nematic phase was recently unveiled using both numerical simulations \cite{Kolli2014a,Kolli2014b,Kolli2016} and density functional theory \cite{Dussi2015,Frezza2014,Cinacchi2017}. In this phase, henceforth denoted as screw-like \cite{Kolli2014a,Kolli2014b,Kolli2016}, the secondary director $\widehat{\mathbf{C}}$ rotates perpendicularly to the main director $\widehat{\mathbf{N}}$ (defined as the eigenvector of $\Lambda_{u_1}$) in a helical fashion as a screw in a cork (see Figure \ref{fig:fig2b}). It is important to stress that this chiral nematic phase is specific for helical particles and is different from the cholesteric phase, common to general chiral particles, where it is the main director $\widehat{\mathbf{N}}$ that revolves with its tip forming a helix around an axis  $\widehat{\mathbf{H}}$ that is perpendicular to the plane of the original direction of $\widehat{\mathbf{N}}$ (see Figure \ref{fig:fig2c}). One of the interesting outcomes of the analysis of the screw-like nematic phase for hard helices stems from the coupling between translation of the helix along the main director $\widehat{\mathbf{N}}$ and a rotation of the secondary director $\widehat{\mathbf{C}}$ as depicted in Figure \ref{fig:fig2d}. This coupling has been observed in both simulations \cite{Kolli2014a,Kolli2014b,Frezza2014,Kolli2016,Cinacchi2017} and in experimental studies of helical flagella \cite{Barry2006,Yardimci2023}. 

In order to identify this particular screw-like phase, it is convenient to introduce a new Veilliard-Baron tensor 
\begin{eqnarray}
  \label{sec1:eq3b}
  \mathbf{Q}_{\widehat{\mathbf{w}}} &=& \left \langle \frac{1}{N} \sum_{j=1}^{N} \left[\frac{3}{2} \widehat{\mathbf{w}}_{j} \widehat{\mathbf{w}}_{j} -\frac{1}{2} \mathbf{I} \right]  \right \rangle
\end{eqnarray}
%\begin{eqnarray}
%  \label{sec1:eq3b}
%  \mathbf{Q}_{\widehat{\mathbf{w}}} &=& \left \langle \frac{1}{N_L} %\sum_{\alpha=1}^{N_{\alpha}} \frac{1}{N_{\alpha}} \sum_{i=1}^{N_{\alpha}} %\left[\frac{3}{2} \widehat{\mathbf{w}}_{i}^{\alpha} %\widehat{\mathbf{w}}_{i}^{\alpha} -\frac{1}{2} \mathbf{I} \right]  \right \rangle
%\end{eqnarray}
which is the counterpart of $\mathbf{Q}_{\widehat{\mathbf{u}}}$ given in Eq.(\ref{sec1:eq3a}).  
Again for the two-fold degeneracy of the uniaxial symmetry of the nematic phase, there will be two positive identical eigenvalues $\Lambda_{w_1}=\Lambda_{w_2}$ and a third negative eigenvalue $\Lambda_{w3}$ due to the traceless nature of the tensor. Hence, the screw-like phase is characterized by a breaking of the uniaxial symmetry and the onset of a characteristic secondary director $\widehat{\mathbf{C}}$ along which all secondary axes $\widehat{\mathbf{w}}$ of the helices preferentially align. This, in turn, is signalled by a removal of the degeneracy ($\Lambda_{w_1}\ne \Lambda_{w_2}$) which can be used as a order parameter for the onset of the screw-like phase. 

An alternative, already used in past work \cite{Kolli2014a,Kolli2014b}, is given by the screw-order parameter $\left \langle P_{1c} \right \rangle$ 
(see \ssref{supp-sec:additional} for this and additional order parameters).
Here, the basic idea is that if all the $\widehat{\mathbf{w}}_i$ align along the secondary director $\widehat{\mathbf{C}}$ then $\langle P_{1c}\rangle \approx 1$, whereas $\langle P_{1c}\rangle \approx 0$ for randomly oriented $\widehat{\mathbf{w}}_i$. In the present formulation, we have implemented an additional layering procedure that avoids the artificial unscrew procedure used in past work \cite{Kolli2014a,Kolli2014b,Kolli2016}, but it has its own shortcomings. The main one is that it is very sensitive to the choice of the width of the layering.

A global parameter $\left \langle \tau_{1} \right \rangle$ can also be implemented to identify the onset of the smectic phase \textrm{SmA} as defined in Ref. \citenum{Bolhuis1997} (see also \ssref{supp-sec:additional}).
Again $\left \langle \tau_{1} \right \rangle \approx 1$ in the smectic phase and $\left \langle \tau_{1} \right \rangle \approx 0$ otherwise. While very useful, it requires preliminary identification of an optimal width, which makes its implementation very time consuming when analysing many different cases.

Finally, the \textrm{SmA}-\textrm{SmB} transition can be identified by the use of the hexatic order parameter $\left \langle \psi_{6} \right \rangle$ \cite{Lopes2021} (see also \ssref{supp-sec:additional}).  
 With this definition, $\langle \psi_{6} \rangle \approx 1$ for hexagonal in-plane ordering and  $\langle \psi_{6} \rangle \approx 0$ otherwise. A bead is defined to be a nearest-neighbor of a bead belonging to another helix if the distance between the two beads is within the first minimum of the perpendicular correlation function (defined further below). As this value clearly depends (albeit not very strongly) on the specific considered state point (i.e. temperature and pressure or volume fraction), even this calculation is not free from ambiguities. 

In addition to global order parameters, local order parameters in the form of suitably defined correlation functions can also be exploited to pin down the exact location of the transition lines. We briefly recall them here and refer to recent past literature for further details (see e.g. Ref. \onlinecite{Lopes2021} and references therein).

A general tendency to form local order can be inferred from the radial distribution function
\begin{eqnarray}
  \label{sec1:eq7}
  g\left(r\right) &=& \frac{1}{4\pi r^2 N } \left \langle \frac{1}{\rho}\sum_{i\ne j} \delta\left(r-r_{ij}\right) \right \rangle
\end{eqnarray}
where $r_{ij}= \vert \mathbf{r}_{ij} \vert = \vert \mathbf{r}_{j}-\mathbf{r}_{i} \vert$. In the case of an anisotropic object such as the helices presented in this study, two additional correlation functions prove particularly useful to identify liquid crystal phases.
The first is the perpendicular correlation function 
\begin{eqnarray}
  g_{\perp}\left(r_{\perp}\right) &=& \frac{1}{2\pi L r_{\perp} N } \left \langle \frac{1}{\rho}\sum_{i\ne j} \delta\left(r_{\perp}-r_{ij}^{\perp}\right) \right \rangle
\label{sec1:eq8}
\end{eqnarray}
where $ \mathbf{r}_{\perp}$ is the projection of $\mathbf{r}$ in the plane perpendicular to the nematic director $\widehat{\mathbf{N}}$, and $r_{\perp}$ is its magnitude.
An oscillatory behavior of $g_{\perp}\left(r_{\perp}\right)$ with specifically located peak points indicates hexagonal orientational order.

The second useful function is the parallel correlation function 
\begin{eqnarray}
  \label{sec1:eq9}
  g_{\parallel}\left(r_{\parallel}\right) &=& \frac{1}{2L^2N} \left \langle  \frac{1}{\rho} \sum_{i\ne j} \delta \left(r_{\parallel}-r_{ij}^{\parallel}\right) \right \rangle
\end{eqnarray}
where $\mathbf{r}_{\parallel} = (\mathbf{r} \cdot \widehat{\mathbf{N}})\widehat{\mathbf{N}}$ and $r_{\parallel}$ is its magnitude. An oscillatory behavior of the $g_{\parallel}\left(r_{\parallel}\right)$ is indicative of the presence of \textrm{Sm}$_{A}$ that becomes  \textrm{Sm}$_{B}$ in the presence of a simultaneous oscillatory behavior of  $g_{\perp}\left(r_{\perp}\right)$. The peak distance in $g_{\parallel}\left(r_{\parallel}\right)$ is a measure of the distance between layers.
Note that while $g(r)$ is normalized in such a way that it tends to $1$ as $r/D \to \infty$, $g_{\perp}\left(r_{\perp}\right)$ and $g_{\parallel}\left(r_{\parallel}\right)$ are normalized such that they tend to $0$ as $r_{\perp}/D  \to \infty$ and $r_{\parallel}/D \to \infty$.  

%A more reliable tool hinges on the use of the parallel correlation function %outlined below. Likewise, the onset of a \textrm{Sm}$_{B}$ phase that includes %hexagonal order in a plane perpendicular to the main nematic director  %\widehat{\mathbf{N}}$ (and including the helix orientational director %$\widehat{\mathbf{C}}$) has its own order parameter that also requires some fine %tuning, and can be conveniently replaced by the perpendicular correlation function %defined below (see e.g. Ref. \onlinecite{Lopes2021} and references therein).
 
The screw-correlation function has also proven useful in this framework. It is defined as
\begin{eqnarray}
\label{sec1:eq10}
g_{1w}(r_{\|}) &=& 
\left \langle \frac{\sum_{i\ne j} \delta(r_{\|}-\mathbf{r}_{ij}\cdot \hat{\mathbf{N}}) (\widehat{\mathbf{w}}_i \cdot \widehat{\mathbf{w}}_j)} 
{\sum_{i\ne j} \delta(r_{\|}-\mathbf{r}_{ij}\cdot \hat{\mathbf{N}})} \right  \rangle \nonumber \\
\end{eqnarray}
and it highlights the tendency of the helices to have their secondary axes  $\widehat{\mathbf{w}}$ aligned. 
%\begin{eqnarray}
%\label{sec1:eq11}
%g_{1u}(r_{\|}) &=& 
%\left \langle \frac{\sum_{i\ne j} \delta(r_{\|}-\mathbf{r}_{ij}\cdot %\hat{\mathbf{N}}) (\widehat{\mathbf{u}}_i \cdot \widehat{\mathbf{u}}_j)} %{\sum_{i\ne j} \delta(r_{\|}-\mathbf{r}_{ij}\cdot \hat{\mathbf{N}})} \right  %\rangle \nonumber \\
%\end{eqnarray}
%plays the same role for the chiral nematic (cholesteric) phase. 

For hard helices it has been argued, based on numerical evidence, that the nematic phase can be either  screw-nematic \textrm{N}$_{S}$ or chiral nematic (cholesteric) \textrm{N}, and not both \cite{Cinacchi2017}. Whether this will also be true for the Janus helices remains to be seen, and for this reason we shall use the general terminology of nematic \textrm{N} for indicating a phase that is \textit{not} screw-nematic, and which is associated with  structureless behavior of $g_{1,\|}^{\widehat{\mathbf{w}}}(r_{\|})$. Note that here and above the short-hand notation $ \sum_{i\ne j} \equiv \sum_{i=1}^N \sum_{\overset{j=1}{j\ne i}}^N$ has been used.
%Note that the following short-hand notation has been used above
%\begin{eqnarray}
%  \label{sec1:eq10}
%  \sum_{i\ne j} \equiv \sum_{i=1}^N \sum_{\overset{j=1}{j\ne i}}^N
%\end{eqnarray}

In general, we found the use of the global screw-nematic order parameter $\langle P_{1c} \rangle $, the smectic order parameter $\langle \tau_1 \rangle$, and the hexatic order parameter $\langle \psi_6 \rangle$ (see \ssref{supp-sec:additional}) to be less informative than the local correlation functions, so they were all monitored but not reported here.
%%%%%%%%%%%%%%%%%%%%%%%%%%%%%%%%%%%%%%%%%%%%%%%%%%%%%%%%%%%%%%%%
\subsection{Diffusion coefficients}
\label{subsec:diffusion}
%%%%%%%%%%%%%%%%%%%%%%%%%%%%%%%%%%%%%%%%%%%%%%%%%%%%%%%%%%%%%%%%
Diffusion was also monitored via simulations. The common definition of the diffusion coefficient $D$ is:
\begin{eqnarray}
\label{sec1:eq10a}
{\cal D} &=& \lim_{t \to \infty} \frac{1}{6t} \left \langle R^2\left(t\right) \right \rangle 
\end{eqnarray}
where the factor $6$ stems from having $2$ possible directions in each of $3$ dimensions. In systems where there is a preferred direction, it is useful to monitor the parallel and perpendicular diffusion coefficients
\begin{eqnarray}
\label{sec1:eq10b}
{\cal D}_{\parallel} &=& \lim_{t \to \infty} \frac{1}{2t} \left \langle R_{\parallel}^2\left(t\right) \right \rangle \\
\label{sec1:eq10c}
{\cal D}_{\perp} &=& \lim_{t \to \infty} \frac{1}{4t} \left \langle R_{\perp}^2\left(t\right) \right \rangle 
\end{eqnarray}
Here, $\langle R^2(t) \rangle$ is defined as the mean-square displacement of the center of mass of the helix at time $t$ with respect to the initial time $t=0$, and $\langle R_{\parallel}^2(t) \rangle$ and $\langle R_{\perp}^2(t) \rangle$ are their counterpart projections along the parallel (to $\widehat{\mathbf{N}}$) and perpendicular directions, respectively.  Notice that this initial time is the initial monitoring time \textit{after} equilibration, unless stated otherwise.
%%%%%%%%%%%%%%%%%%%%%%%%%%%%%%%%%%%%%%%%
\subsection{Virial expansion and Boyle temperature}
\label{subsec:virial}
%%%%%%%%%%%%%%%%%%%%%%%%%%%%%%%%%%%%%
The virial expansion  for pressure $P$ is given by
\begin{eqnarray}
  \label{sec1:eq11a}
\frac{\beta P}{\rho} &=& 1+ \rho B_2\left(T\right) + \rho^2 B_3\left(T\right) + \ldots  
\end{eqnarray}
The Boyle temperature $T_B$ is defined by the condition $B_2(T_B)=0$ \cite{Hansen2006}. For $T>T_B$, $B_2(T)>0$ and repulsions dominate; For $T<T_B$, $B_2(T)<0$ and attractions dominate. Hence $T_B$ represents the transition temperature between entropically and energetically dominated regimes.
%%%%%%%%%%%%%%%%%%%%%%%%%%%%%%%%%%%%%%%%%%%%%%%%%%%%%%%%%%%%%%%%%%%%%
\section{From hard to fully attractive helices}
\label{sec:results}
%%%%%%%%%%%%%%%%%%%%%%%%%%%%%%%%%%%%%%%%%%%%%%%%%%%%%%%%%%%%%%%%%%%%

We have carried out ramps in both pressure (for NPT-MC simulations) and volume fraction (for NVT-MD simulations). In the first case, we considered 10 different reduced pressures ranging from $0.2$ to $2.0$ for each of the $N=2400$ NPT-MC simulations and ranging from $0.1$ to $1.0$ for the $N=972$ NPT-MC simulations. In the second case, we considered 12 different volume fractions from $\eta=0.1$ to $\eta=0.65$. In both cases, the simulations were repeated for three different (reduced) temperatures $T^{*}=1.0,0.5,0.1$, and four different coverages, from the case of hard helices to the case of fully attractive helices, passing through the single attractive bead case (out of 15) and the Janus limit of approximately half attractive beads. The three temperatures were selected on the basis of the analysis of the second virial coefficients reported below.
%%%%%%%%%%%%%%%%%%%%%%%%%%%%%%%%%%%%%%%%%%%%%%%%%%%%%%%%%%%%%%%%%%%%5
\subsection{Second virial coefficient $B_2$ and Boyle temperature $T_B^{*}$ at different coverages.}
\label{subsec:B2_chi}
%%%%%%%%%%%%%%%%%%%%%%%%%%%%%%%%%%%%%%%%%%%%%%%%%%%%%%%%%%%%%%%
It proves convenient to compute the reduced second virial coefficient black $B_2^{*}=B_2/D^3$ as a function of the reduced temperature  $T^{*}$, as well as the reduced Boyle temperature $T_{B}^{*}$, where $B_2^{*}(T_{B}^{*})=0$, as a function of the coverage $\chi$. To this aim, we have followed  past suggestions \cite{Yethiraj1991,Heyes2015} and recall the definition of the second virial coefficient
\begin{eqnarray}
\label{sec2:eq1}
B_2\left(T\right) &=& -\frac{1}{2} \int d \mathbf{r} \left \langle f_{12} \left(\mathbf{r},\widehat{\mathbf{u}}_1, \widehat{\mathbf{u}}_2 \right) \right \rangle_{\widehat{\mathbf{u}}_1, \widehat{\mathbf{u}}_2}
\end{eqnarray}
where we have introduced the Mayer function \cite{Hansen2006}
\begin{eqnarray}
\label{sec2:eq2}
f_{12} \left(\mathbf{r},\widehat{\mathbf{u}}_1, \widehat{\mathbf{u}}_2 \right) &=& e^{-\beta \phi\left(\mathbf{r},\widehat{\mathbf{u}}_1, \widehat{\mathbf{u}}_2\right)} -1
\end{eqnarray}
as well as the average over the angular orientations $\widehat{\mathbf{u}}_1, \widehat{\mathbf{u}}_2$ of the two helices
\begin{eqnarray}
\label{sec2:eq3}
\left \langle \ldots \right \rangle_{\widehat{\mathbf{u}}} &=& \frac{1}{4\pi} \int d \widehat{\mathbf{u}} ~ \ldots
\end{eqnarray}
Here $\phi(\mathbf{r}, \widehat{\mathbf{u}}_1, \widehat{\mathbf{u}}_2)$ is the sum of all potentials between two sites on different helices, $\mathbf{r}$ being the displacement between the center of mass of the two helices and $\widehat{\mathbf{u}}_1$ and $\widehat{\mathbf{u}}_2$ defining their orientations in space. 
This integration was performed numerically by generating a large number $N_c$ of independent configurations of two identical helices in a cubic box of volume $V$.
\color{black}
$B_2$ can then be computed as \cite{Yethiraj1991,Munao2015}
\begin{eqnarray}
\label{sec2:eq5}
B_2\left(T\right) &=& - \frac{V}{2 N_c} \left \langle f_{12} \right \rangle_{\mathbf{r},\widehat{\mathbf{u}}_1, \widehat{\mathbf{u}}_2}
\end{eqnarray}
\color{black}
Note that the average over all possible $\mathbf{r}$ has been included in the last average appearing in Eq.(\ref{sec2:eq5}).

For calculation speed, note that the Mayer function can assume three different types of values depending on the relative position $\mathbf{r}_{12} \equiv \mathbf{r}$ of the two closest beads between the two helices:
\color{black}
\begin{equation}
\label{sec2:eq4}
 f_{12}\left(\mathbf{r}, \widehat{\mathbf{u}}_1, \widehat{\mathbf{u}}_2 \right) =
 \begin{cases}
 -1 & \mbox{if helices overlap i.e.} ~r < D\\
 e^{n_c/n_s^2T^{*}}-1 & \mbox{if no overlap and } D < r < R_c\\
 0 &  r > R_c
 \end{cases}
\end{equation}
\color{black}
where $r$ is the smallest distance between two spheres on different helices, and $n_c$ is the number of square-well contacts ($D< r < R_c$ between attractive spheres on two different helices). 

In the case of a hard overlap, the Mayer function can be immediately evaluated, with $\phi \to \infty$ forcing the value of the corresponding Mayer function to $-1$, which we then add to the accumulated average.
In the case of no overlaps, but at least one interaction, care must however be exercised in doing this calculation in the presence of multiple sources of attraction. As $-\epsilon/n_s^2$ is the energy contribution from each favourable interaction (see Eq.(\ref{sec1:eq2})), the energy must first be summed with a double loop over the beads belonging to the two helices, adding the $-\epsilon/n_s^2$ energy contribution from each favourable interaction, before evaluating the Mayer function.

By gradually varying the temperature $T$ until $B_2$ vanishes, the Boyle temperature $T_B$ was obtained. 

We explicitly checked that the limiting behaviour at high temperatures is consistent with Onsager's celebrated limit for hard spherocylinders \cite{Onsager1949,Heyes2015}
\begin{eqnarray}
\label{sec2:eq6}
B_2 &=& \pi \left(\frac{2D^3}{3} + L D^2 + \frac{L^2 D}{4} \right)
\end{eqnarray}
%%%%%%%%%%%%%%%%%%%%%%%%%%%%%%% Fig 3 %%%%%%%%%%%%%
\begin{figure}[htpb]
  \centering
  \captionsetup{justification=raggedright,width=\linewidth}
  \begin{subfigure}[b]{\linewidth}
   \includegraphics[trim=10 10 20 20,clip,width=\linewidth]{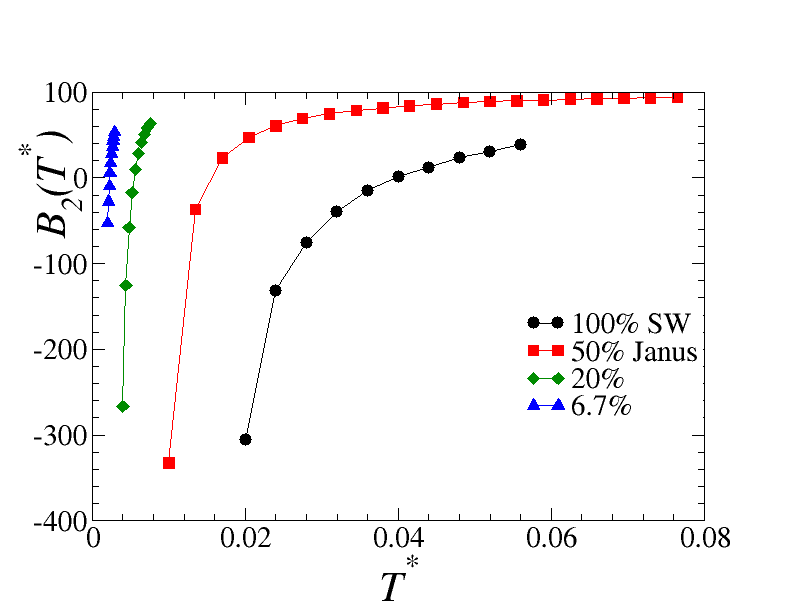}
   \caption{}\label{fig:fig3a}
  \end{subfigure}%
  \hfill
    \begin{subfigure}[b]{\linewidth}
   \includegraphics[trim=20 20 20 20,clip,width=\linewidth]{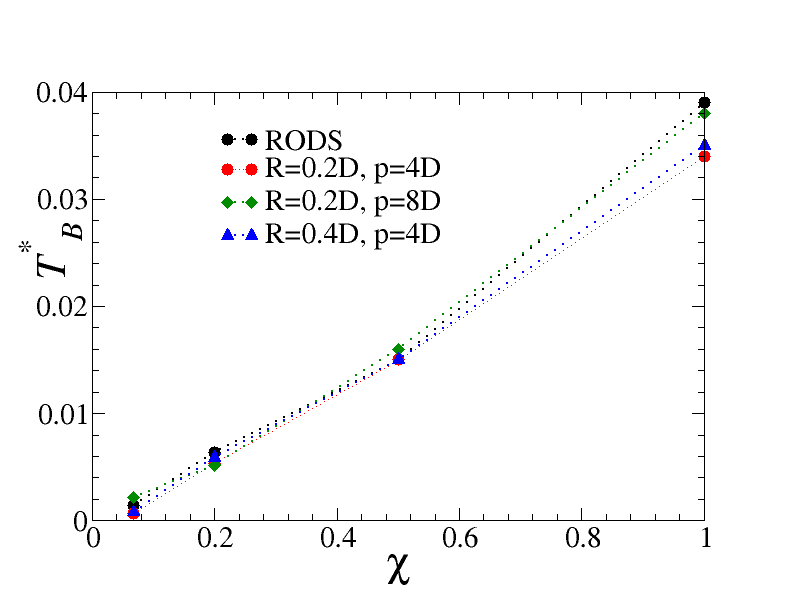}
   \caption{}\label{fig:fig3b}
  \end{subfigure}
  \caption{(a) The second virial coefficient $B_2$ as a function of the reduced temperatures $T^{*}$ at different coverages $\chi$  for $R=0.2D$ and $p=8D$; (b) The reduced Boyle temperature $T_B^{*}$ as a function of the coverage $\chi$ for the different shapes of the helix. The curve labelled as "RODS" depicts the Boyle temperature derived by the vanishing of $B_2$ for rods with length $L/D=10$.}
  \label{fig:fig3}
\end{figure}
%%%%%%%%%%%%%%%%%%%%%%%%%%%%%%%%%%%%%%%%%%%%%%%%%%%%%%%%%
As a representative calculation, Figure \ref{fig:fig3a} shows the results of the reduced second virial coefficient $B_2^{*}$  as a function of the reduced temperature $T^{*}$ for helices with $R=0.2D$ $p=8D$. Figure \ref{fig:fig3b} reports the corresponding Boyle temperature as a function of the coverage $\chi$ for all of the helical shapes considered in this study. A comparison with the straight rod limit $p \to \infty$ is also displayed. \stref{supp-tab:boyle} reports the numerical values of the Boyle temperatures in the case of helices with $R=0.2D$ and $p=8D$.

Two main insights can be extracted from this analysis. First, there appear to be no major differences between the three shapes and the straight rod counterpart at this first approximate level, thus justifying our choice of focusing on one specific shape. Second, the Boyle temperature varies from $T_B^{*} \approx 0.04$ for the fully attractive SW case to $T_B^{*} \approx 0.0015$ in the $6.7\%$ limit where only one bead is attractive. As the Boyle temperature signals the transition from an entropy-dominated regime ($T^{*} \gg T_B^{*}$) to an energy-dominated one  ($T^{*} \ll T_B^{*}$), this means that significantly low temperatures are necessary to observe competition between the liquid crystal phases observed in the purely repulsive hard helices case \cite{Frezza2013,Kolli2014a,Kolli2014b,Frezza2014,Cinacchi2017}, and a possible gas-liquid transition akin to that observed in the spherical Janus counterpart \cite{Sciortino2009,Sciortino2010}. In the present study, this low $T^{*}$ regime will not be addressed as its analysis will be the subject of a companion paper on Janus rods \cite{Wood2023}. Rather, we will focus our attention on the effect of a relatively weak attraction to the liquid-crystal phase behaviour by probing temperatures in the range $0.1 \le T^{*} \le 1.0 $. As  $T^{*}=0.1$ is larger then $T_B$ by a factor $\approx 2$ for fully attractive helices but by a factor $\approx 100$ for helices with a single attractive site ($\approx 6.7\%$ case), a correct interpretation of the results must be drawn accordingly. 

%%%%%%%%%%%%%%%%%%%%%%%%%%%%%%%%%%%%%%%%%%%%%%%%%%%%%%%%%%%%%%
\subsection{The hard helices case}
\label{subsec:hard}
%%%%%%%%%%%%%%%%%%%%%%%%%%%%%%%%%%%%%%%%%%%%%%%%%%%%%%%%%%%%%%%
Before discussing the case of attractive helices, we now briefly recall the hard helices case discussed in previous work \cite{Kolli2014a,Kolli2014a,Kolli2016} with a two fold aim. First, it will allow a quantitative comparison in terms of finite size within the same  isobaric-isothermal NPT-MC simulations, from $N=972$ of past work to $N=2400$ of the present study. Second, it will allow to test for consistency by a comparison with isochoric-isothermal NVT-MD calculations using $N=4068$ helices. In all cases, the standard initial condition was taken as a system of initially parallel helices assembled at different densities. Supplementary Figure SI shows the approach to equilibrium for both NPT-MC and NVT-MD simulations. This shows that the NPT-MC simulations have difficulty reaching complete equilibration at the highest considered pressures despite the relatively long equilibration times ($5 \times 10^6$ MC steps). In contrast, the NVT-MD simulations appear unaffected by this shortcoming, and so results on partially attractive helices mainly rely on them. It is reassuring, however, that the two sets of simulations provide compatible results when both are equilibrated.

%%%%%%%%%%%%%%%%%%%%%%%%%%%%%% FIG 4 %%%%%%%%%%%%%%%%%%%%%%%%%%%%%%%%%%%%%%%%%%%
\begin{figure}[htbp]
 \centering
  \captionsetup{justification=raggedright,width=\linewidth}
  \begin{subfigure}[b]{\linewidth}
   \includegraphics[trim= 20 20 50 50,clip,width=\linewidth]{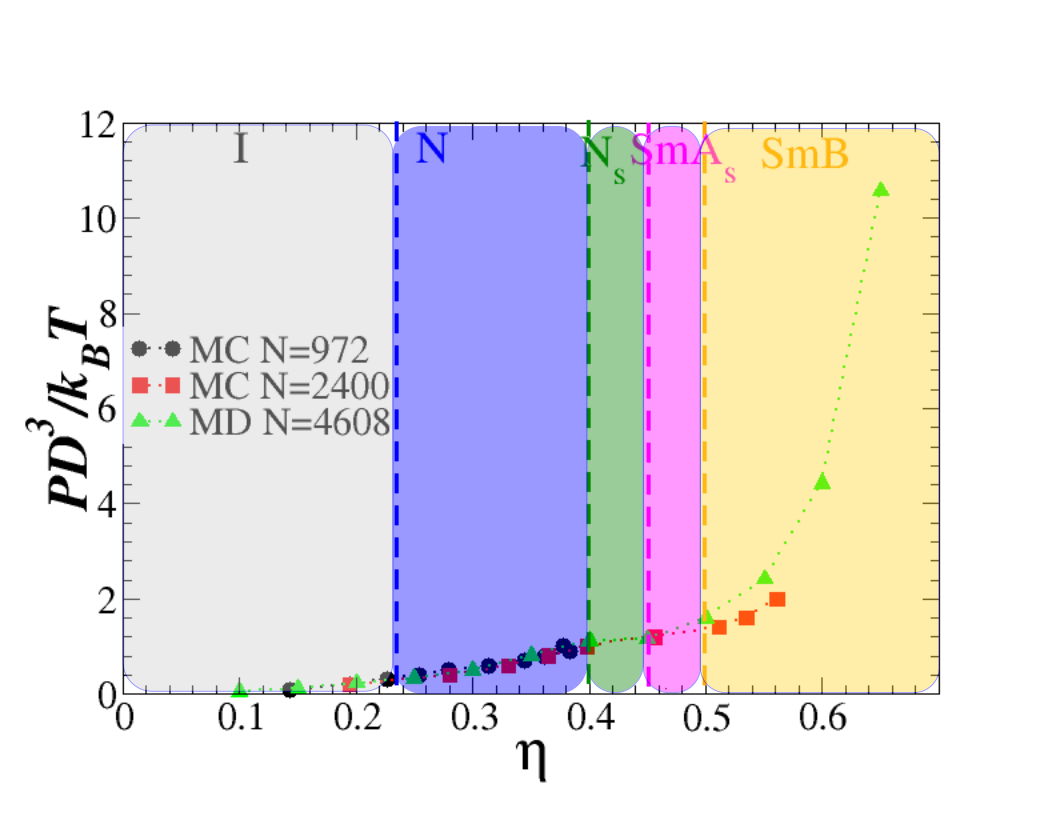}
   \caption{}\label{fig:fig4a}
\end{subfigure}
 \begin{subfigure}[b]{\linewidth}
 \centering
  \captionsetup{justification=raggedright,width=\linewidth}
   \includegraphics[trim= 0 0 0 0,clip,width=\linewidth]{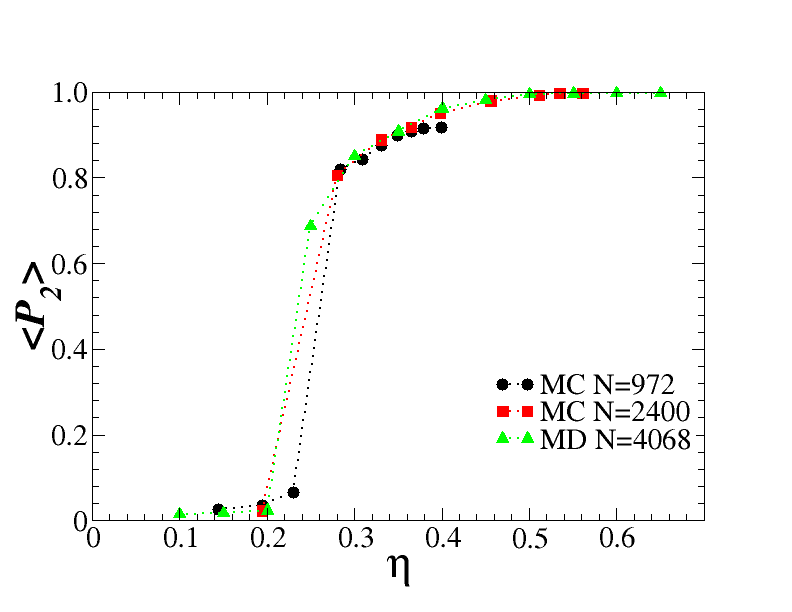}
   \caption{}\label{fig:fig4b}
 \end{subfigure}
 \caption{(a) Equation of state $P D^3 / k_B T$ as a function of the packing fraction $\eta$ for the case of hard helices. The three sets of points refer to Monte Carlo simulations with $N=972$ helices (black circle) and $N=2400$ (red squares), and to MD simulations with $N=4068$ (green triangles). Parallel dotted lines indicate the transitions identified in past work ( see e.g. \cite{Cinacchi2017}) and refer to the \textrm{I}-\textrm{N} isotropic-nematic transition (blue line),  \textrm{N}-\textrm{N}$_{s}$ nematic-screw-nematic transition (green line), \textrm{N}$_{s}$-\textrm{SmA}$_{\text{s}}$ screw-nematic-screw-smectic A transition (magenta line),  \textrm{SmA}$_{\text{s}}$-\textrm{SmB} screw-smectic A-polar-smectic B transition (orange line); (b) The $\langle P_2 \rangle$ order parameter as a function of the packing fraction $\eta$ for the same three sets of simulations as in part (a).}
   \label{fig:fig4}
   \end{figure}
   %%%%%%%%%%%%%%%%%%%%%%%%%%%%%%%%%%%%%%%%%%%%%%%%%%%%%%%%%%%%%%%%

Figure \ref{fig:fig4a} reports the equation of state (reduced pressure $P D^3 / k_B T$ versus packing fraction $\eta$) from both NVT-MD with $N=4068$ helices (green triangle)  and NPT-MC calculations with $N=2400$ (red squares), and contrasted with the results of the smaller size NPT-MC calculations with $N=972$ (black spheres) used in previous studies \cite{Kolli2014a,Kolli2014b,Kolli2016}). Color coded vertical lines indicate the putative transitions from isotropic \textrm{I} to nematic \textrm{N} (blue line), nematic \textrm{N} to screw-nematic \textrm{N}$_{s}$ (green line), screw-nematic \textrm{N}$_{s}$ to screw-smectic A \textrm{SmA}$_{s}$ (magenta line), screw-smectic A \textrm{SmA}$_{s}$ to polar-smectic B \textrm{SmB} (orange line) as identified in previous work \cite{Kolli2014a,Kolli2014b,Kolli2016,Cinacchi2017}. Our present estimates confirm these findings. As anticipated, the  isotropic \textrm{I} to nematic \textrm{N} (blue line) can be located by considering the nematic order parameter $\langle P_2 \rangle$ that also coincides with the maximum eigenvalue $\Lambda_{u_1}$ of $\mathbf{Q}_{\widehat{\mathbf{u}}}$ (Eq.\ref{sec1:eq3a}). This is shown in Fig. \ref{fig:fig4b} for both NVT-MD  and NPT-MC calculations, updated and original, as in Figure \ref{fig:fig4a}. The upswing of the curve at $\eta \approx 0.25$ marks the end of the isotropic phase and the onset of the nematic one.  

In Supplementary Figure SV we show results for the equation of state from both NPT-MC and NVT-MD simulations at different temperatures. The closest match between the simulated state points appears to be when $T^{*}=1.0$ as one could have expected from the outset. We will then use this temperature value when contrasting the two different techniques hereafter.

Supplementary Figure SVII and SVIII report the equilibrium behaviour of the three eigenvalues $\Lambda_{u_{i}}$ ($i=1,2,3$) of $\mathbf{Q}_{\widehat{\mathbf{u}}}$ (Eq.\ref{sec1:eq3a}) as a function of $\eta$, as obtained from the NPT-MC and NVT-MD calculations. This shows that the maximum eigenvalue  $\Lambda_{u_{1}}$ progressively increases from $\approx 0$ for $\eta$ small to $\approx 1$ for large $\eta$. Accordingly, the other two eigenvalues  $\Lambda_{u_{2}}$ and $\Lambda_{u_{3}}$ also move from  $\approx 0$ for $\eta$ small to $\approx -0.5$ for large $\eta$. Their equality is an indication of the uniaxial symmetry of the system at this stage.

%%%%%%%%%%%%%%%%%%%%%% FIG 5 %%%%%%%%%%%%%%%%%%%%%%%%%%%%%%%%%
\begin{figure}
    \centering
    \includegraphics[trim=0 0 0 0,clip,width=1.0\linewidth]{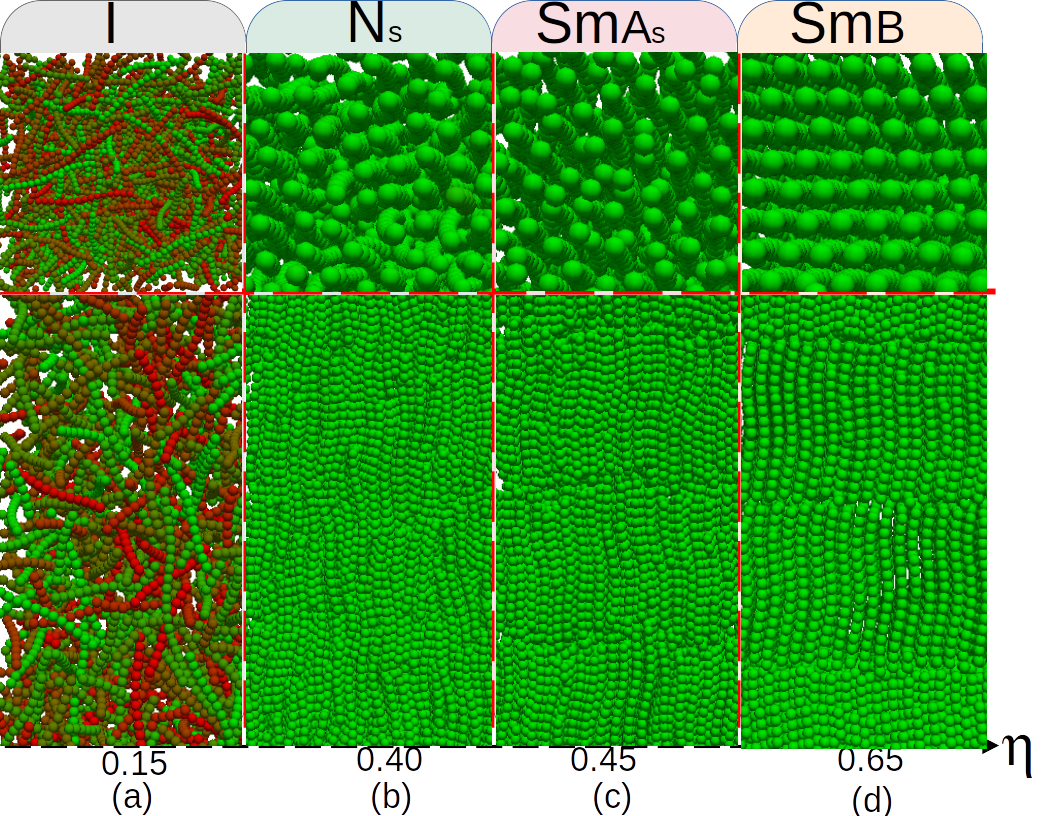}
    \caption{Representative snapshots (top and side views) of state points appearing in Figure \ref{fig:fig4} as obtained by MD $N=4068$. Different helices are color-coded according to the orientation $\widehat{\mathbf{u}}$ of the helix main axis, and both top and side views are depicted in all cases, with tick red dotted lines separating different phases. (a) \textrm{I} state point $\eta=0.15$; (b) \textrm{N}$_{s}$ state point $\eta=0.40$; (c) \textrm{SmA}$_{s}$ state point $\eta=0.45$; (d) \textrm{SmB} state point $\eta=0.65$. Screw phases are not easily identifiable here as they are best illustrated in Figure \ref{fig:fig6}. The light colored backgrounds that appear behind the phase identification letters are consistent with the phases in Figure \ref{fig:fig4a} and thick red dashed lines separate different phases. Visualizations here and below were done using the Ovito Package \cite{Stukowski2010}.}
    \label{fig:fig5}
\end{figure}
%%%%%%%%%%%%%%%%%%%%%%%%%%%%%%%%%%%%%%%%%%%%%%%%%%%%%%%%%%%%%%%%%
%%%%%%%%%%%%%%%%%%%%%% FIG 6 %%%%%%%%%%%%%%%%%%%%%%%%%%%%%%%%%
\begin{figure}
    \centering
    \includegraphics[trim=20 10 20 20,clip,width=1.0\linewidth]{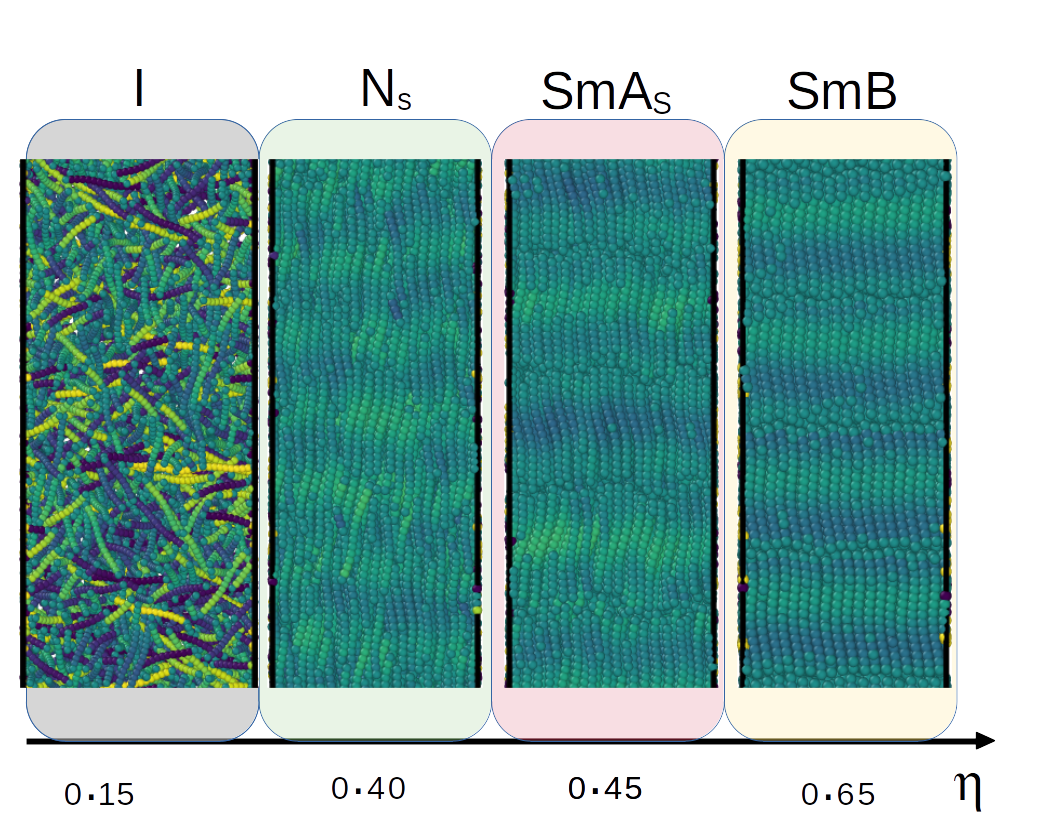}
    \caption{Same as in Figure \ref{fig:fig5} but now with the \textcolor{black}{beads} colored according to the local tangent as in Ref. \cite{Kolli2014a,Kolli2014b}. The color changes as the tangent moves along the helix and thus the periodicity of the color pattern is equal to the pitch of the helix. Supplementary Figure SII shows how the tip of the local tangent performs a conical path along the helix. Note that this modality mirrors the experimental one used in Ref. \cite{Barry2006}. Here only side views are presented and the light colored background is color-coded as in the phase diagram of Figure \ref{fig:fig4a}. The periodic stripes appearing here in the \textrm{N}$_{s}$ and \textrm{SmA}$_{s}$ phases are indicative of screw-like phases. The stripe periodicity of the  screw-nematic \textrm{N}$_{s}$ phase is equal to the pitch $p$ of the helix.}
    \label{fig:fig6}
\end{figure}
%%%%%%%%%%%%%%%%%%%%%%%%%%%%%%%%%%%%%%%%%%%%%%%%%%%%%%%%%%%%%%%%%
%%%%%%%%%%%%%%%%%%%%% FIG 7 %%%%%%%%%%%%%%%%%%%%%%%%%%%%%%%%%
\begin{figure}
    \centering
    \includegraphics[trim=0 0 0 0,clip,width=1.0\linewidth]{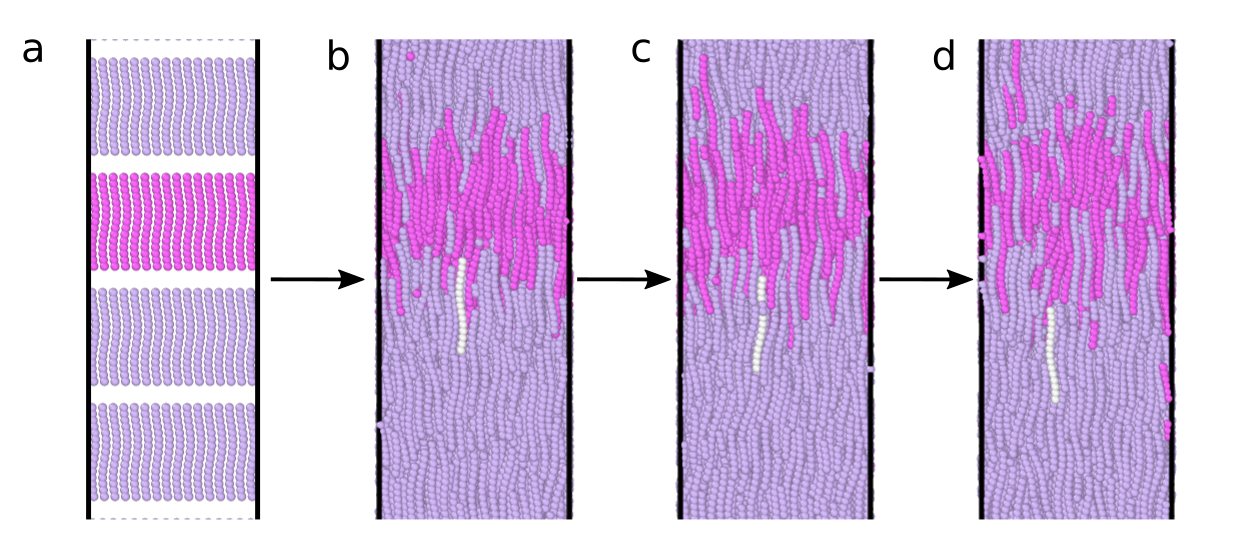}
    \caption{A visual representation of the screw-like mechanism within a screw-nematic phase \textrm{N}$_{s}$ at $\eta=0.40$. Helices colored in dark magenta in a) are originally aligned in a layer and then eventually dispersed within the computational box by the dynamics. Within this framework, the white helix is shown to diffuse down from b) to c) to d) via a translation (along the $\widehat{\mathbf{N}}$ axis) coupled with a rotation (about the main axis  $\widehat{\mathbf{n}}$ of each helix) as a screw in a bolt, in close analogy with the experiments in Refs. \cite{Barry2006,Yardimci2023}.}
    \label{fig:fig7}
\end{figure}
%%%%%%%%%%%%%%%%%%%%%%%%%%%%%%%%%%%%%%%%%%%%%%%%%%%%%%%%%%%%%%%%%
Close-up snapshots of the representative state points are reported in Figure \ref{fig:fig5}, with different helices color-coded according to their directions. Both top and side views are depicted in all cases. The four depicted snapshots refer to states with $\eta=0.15$ (isotropic \textrm{I} phase), $\eta=0.40$ (screw-nematic \textrm{N}$_{s}$ phase), $\eta=0.45$ (\textrm{SmA}$_{s}$  phase), and $\eta=0.65$ (smectic B \textrm{SmB} phase). The presence of the screw-nematic phase is not obvious in Figure \ref{fig:fig5}(b) and can be more easily seen in Figure \ref{fig:fig6}, where the same snapshots have been color-coded according to the local tangent rather than the local director $\widehat{\mathbf{u}}$. Particularly evident are the stripes in the $\eta=0.40$ (screw-nematic \textrm{N}$_{s}$ phase), $\eta=0.45$ (\textrm{SmA}$_{s}$  phase), with the screw-like nature of the \textrm{SmA}$_{s}$ originating from the screw-like nature of the \textrm{N}$_{s}$ phase).  Also the smectic B \textrm{SmB} phase presents a striped pattern, indicating that all $\widehat{\mathbf{w}}$ within a layer are in phase. However, in contrast with its   \textrm{SmA}$_{s}$ counterpart, stripes in different layers are uncorrelated. This was rationalized \cite{Kolli2014a,Kolli2014b} on the basis that the driving force originating the \textrm{SmB} phase stems mainly from the requirement of minimizing excluded volume at the expenses of orientational ordering.  

Further insights on the presence of the screw-nematic \textrm{N}$_{s}$ phase can be seen in Figure \ref{fig:fig7} where a subset of helices initially equilibrated at $\eta=0.40$ and confined within a given layer have been highlighted via a different color (dark magenta) and they can be seen to visibly diffuse up and down along the main director $\widehat{\mathbf{N}}$ (as is also visible in Movie1 reported in Supplementary Information). The three snapshots of Figure \ref{fig:fig7} show one particular helix colored in white that is seen to perform a screw-like motion akin to that found for helical flagellae by \citeauthor{Barry2006}\cite{Barry2006} and more recently by \citeauthor{Yardimci2023} \cite{Yardimci2023}. As helices tend to align due to an increasing concentration, they loose rotational entropy. This can be compensated by corkscrewing up and down along the main director $\widehat{\mathbf{N}}$.

%%%%%%%%%%%%%%%%%%%%%%%%%%%% Fig 8 %%%%%%%%%%%%%
\begin{figure*}[htpb]
  \centering
  \captionsetup{justification=raggedright,width=\linewidth}
  \begin{subfigure}{0.45\textwidth}
   \includegraphics[trim= 10 10 10 10,clip,width=\linewidth]{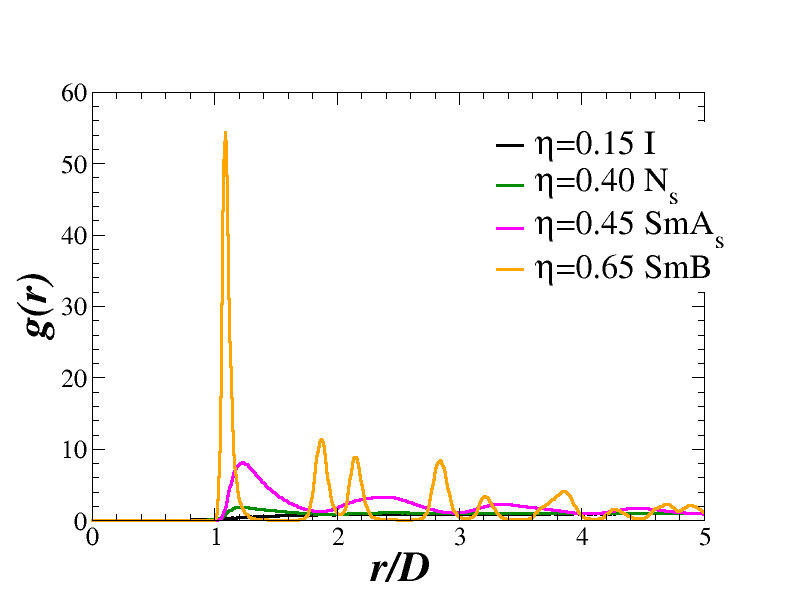}
    \caption{}\label{fig:fig8a}
   \end{subfigure}
   \begin{subfigure}{0.45\textwidth}
   \includegraphics[trim= 0 10 10 10,clip,width=\linewidth]{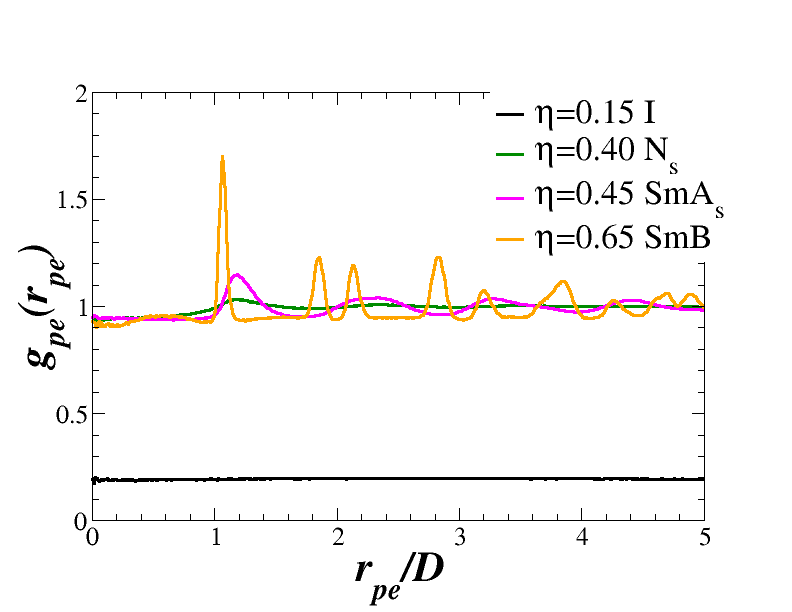}
    \caption{}\label{fig:fig8b}
   \end{subfigure} \\
   \begin{subfigure}{0.45\textwidth}
   \includegraphics[trim= 0 10 10 10,clip,width=\linewidth]{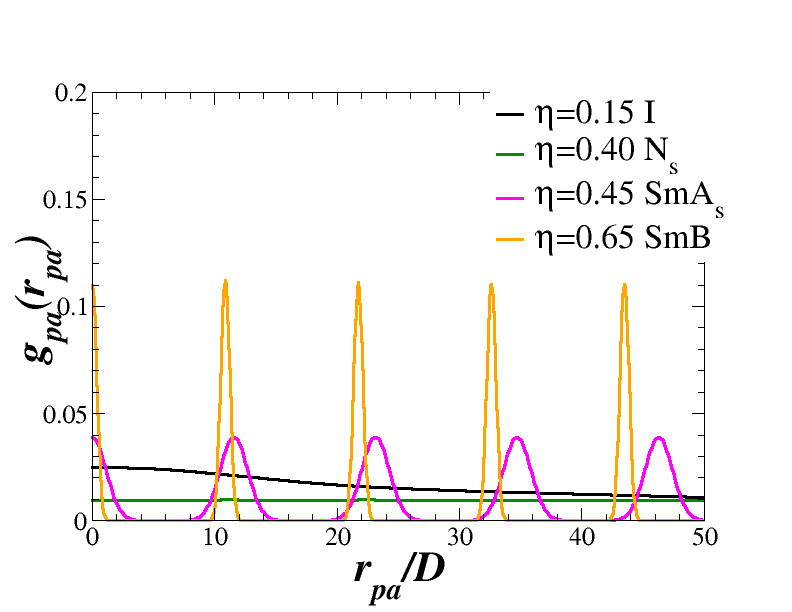}
    \caption{}\label{fig:fig8c}
   \end{subfigure}
   \begin{subfigure}{0.45\textwidth}
   \includegraphics[trim= 10 10 10 10,clip,width=\linewidth]{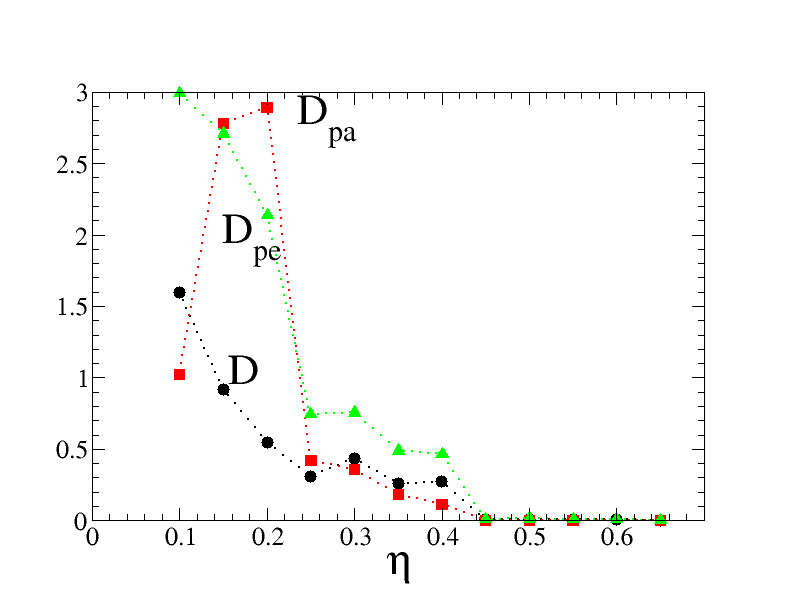}
    \caption{}\label{fig:fig8d}
   \end{subfigure}
   \caption{Fully repulsive helices represented by a WCA model at $T^{*}=1.0$. (a) Radial correlation functions $g(r)$ as a function of the reduced distance $r/D$ between the centers of the helices at different state points; (b) Perpendicular correlation function $g(r_{\text{pe}})$ as a function of the reduced perpendicular distance $r_{\text{pe}}/D$
   where $r_{\text{pe}} \equiv r_{\perp}$; (c) Parallel correlation function $g(r_{\text{pa}})$ as a function of the reduced parallel distance $r_{\text{pa}}/D$ where $r_{\text{pa}} \equiv r_{\parallel}$. Color coding of the different state points here is the same as in Figure \ref{fig:fig4}; (d) Overall diffusion coefficient $D$ as a function of the packing fraction $\eta$, contrasted with its perpendicular $D_{pe} \equiv D_{\perp}$ and parallel $D_{pa} \equiv D_{\parallel}$ components. } 
  \label{fig:fig8}
\end{figure*}
%%%%%%%%%%%%%%%%%%%%%%%%%%%%%%%%%%%%%%%%%%%%%%%%%%%%%%%%%%%%%%

%As discussed in Section \ref{subsec:order}, the onset of a screw-nematic ordering can be %located by looking at the screw-nematic order parameter $\langle P_{1,c} \rangle$ %(Eq.(\ref{sec1:eq4})). A value close to unit of this value indicates that  the helices are %all locked in phase with nearly identical azimuthal angles. 

The exact location of the phase boundaries can be conveniently obtained by considering the correlation functions discussed in Section \ref{subsec:order}. Figure \ref{fig:fig8a} displays the radial distribution function of the center-of-masses of the helices, as a function of $r/D$ for the same state points reported in Figure \ref{fig:fig5}. The color coding of the different curves is the same as the phases reported in Figure \ref{fig:fig4a}. Starting from $\eta \approx 0.4$, the first and second shell peaks begin developing before additional crystal-like peaks appear at $\eta =0.65$. The location of the peaks along the $r$ axis nearly matches those of the perpendicular correlation function $g_{pe}\equiv g_{\perp}(r_{\perp})$ along the $r_{pe} \equiv r_{\perp}$ axis (Figure \ref{fig:fig8b}), thus locating parallel neighbouring helices with an in-plane hexatic ordering characteristic of the \textrm{SmB} phase (see below).
%A complementary local probe of this transition stems from the correlation function %$g_{1,\parallel}^{\widehat{\mathbf{w}}}(r_{\parallel})$ (Eq. \ref{sec1:eq8}).

The onset of smectic phases is signalled by the parallel correlation function $g_{pa} \equiv g_{\parallel}(r_{\parallel})$ (see Figure \ref{fig:fig8c}) that develops marked oscillating from $\eta =0.45$ onward, in agreement with Figure \ref{fig:fig4a}. Note the significant range difference of the $r_{pa} \equiv r_{\parallel}$ axis compared to the case of $r$ (Figure \ref{fig:fig8a}) and $r_{pe}$ (Figure \ref{fig:fig8b}) that is clearly due to the large aspect ratio of the helices that require an asymmetric computational box elongated along the $\mathbf{N}$ axis.  As the helices were originally in a screw-like phase, the obtained phase is a \textrm{SmA}$_{s}$, as indicated by the magenta vertical line of Figure \ref{fig:fig4a}.  At $\eta \approx 0.5$ the system undergoes a \textrm{SmA}$_{s}$ to \textrm{SmB}  transition (see orange line in Fig. \ref{fig:fig4a}), with in-plane hexagonal symmetry combined with alignment of all secondary vectors $\widehat{\mathbf{w}}$ within the plane. Here, however, entropic gain in off-setting parallel alignment between consecutive layers disfavors AAA stacking as well as the screw-like ordering, so that consecutive layers have secondary directors $\widehat{\mathbf{C}}$ (see Fig. \ref{fig:fig2b}) that become uncorrelated, mirroring the loss of AAA alignment for the positional ordering \cite{Kolli2014b}. The in-plane hexagonal ordering is clearly visible in the perpendicular correlation function $g_{\perp} (r_{\perp})$ displayed in Figure \ref{fig:fig8b}, with the well-developed peaks at $\eta=0.65$ showing a characteristic $1:\sqrt{2}$ periodicity.

Another interesting point stems from the analysis of the equilibrium values of the three eigenvalues $\Lambda_{w_{i}}$ ($i=1,2,3$) of $\mathbf{Q}_{\widehat{\mathbf{w}}}$ (Eq.\ref{sec1:eq3b}) as a function of $\eta$. This analysis mirrors the same analysis on the eigenvalues of $\mathbf{Q}_{\widehat{\mathbf{u}}}$ discussed earlier.  This is reported in Supplementary Figure SVII and SVIII for NPT-MC and NVT-MD respectively. Complementary to that analysis, here the two largest eigenvalues $\Lambda_{w_{1,2}}$ are positive $\approx 0.25$ and identical (i.e. degenerate), with the third eigenvalue negative and equal to $\approx -0.5$ for unscrew phases. Once again, the degeneracy of the first two eigenvalues stems from the uniaxial symmetry of the nematic phase and breaks down at the onset of the screw phases, where $\Lambda_{w_{1}} \ne \Lambda_{w_{2}}$ for volume fractions from $\eta \approx 0.40$ to $\eta \approx 0.55$, in good agreement with the phase diagram of Figure \ref{fig:fig4a} and with evidence from the corresponding correlation function previously discussed. It is important to stress that in all cases, the entries of the tensor matrices are averages over different configurations, and hence fluctuations from one calculation to another are certainly possible.

The use of MD simulations also makes it possible to calculate the overall diffusion coefficient $D$ (defined in Eq. \ref{sec1:eq10a}), the parallel diffusion coefficient $D_{\parallel}$ (Eq. \ref{sec1:eq10b}), and the perpendicular diffusion coefficient $D_{\perp}$ (Eq. \ref{sec1:eq10c}). These were calculated after equilibration and are plotted in Figure \ref{fig:fig8d} as a function of the packing fraction $\eta$. As expected on physical grounds, all three diffusion coefficients markedly decrease as the system transitions from the isotropic phase into the nematic phase at $\eta \approx 0.25$. The diffusion coefficients are still significant in the screw-nematic \textrm{N}$_{s}$, as expected from Figure \ref{fig:fig7}), eventually becoming negligibily small as the system enters the smectic phase ($\eta \geq 0.45$). Interestingly, the lateral diffusion appears to be more pronounced than the longitudinal diffusion (i.e. along the main director $\widehat{\mathbf{N}}$) after entering into the nematic phase, whereas it is subdominant just before the \textrm{I}-\textrm{N} transition. No evidence of either intralayer or interlayer diffusion is observed in any of the smectic phases \cite{Cinacchi2009}.

%%%%%%%%%%%%%%%%%%%%%%%%%%%% Fig9 %%%%%%%%%%%%%
\begin{figure}[htpb]
  \centering
  \includegraphics[trim= 0 10 10 10,clip,width=\linewidth]{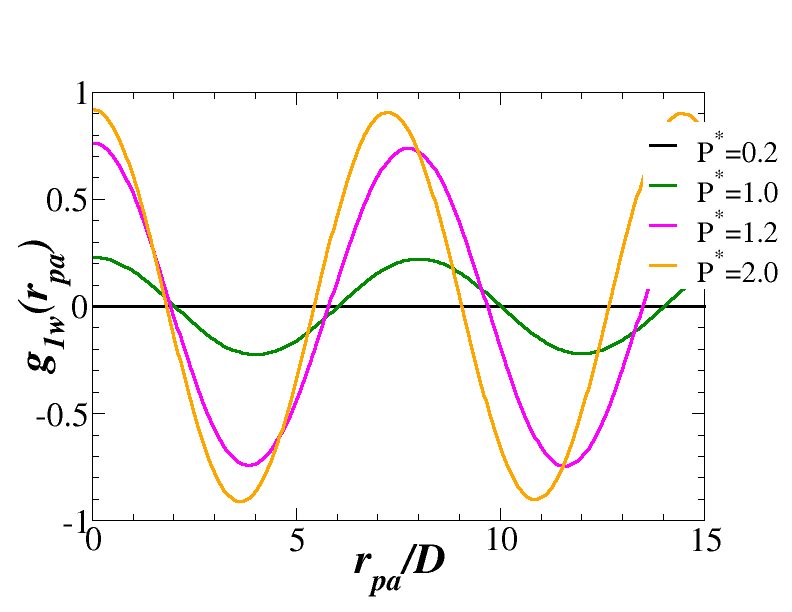}
   \caption{(NPT-MC results with $N=2400$ helices for $g_{1w}(r_{pa})$ as a function of $r_{pa} \equiv r_{\parallel}$ for the same volume fractions as in Figure \ref{fig:fig8}. Note that we have used the same color coding though the mapping between reduced pressures and volume fraction is approximate as inferred from the equation of state (see Figure \ref{fig:fig4a}). } 
  \label{fig:fig9}
\end{figure}
%%%%%%%%%%%%%%%%%%%%%%%%%%%%%%%%%%%%%%%%%%%%%%%%%%%%%%%%%%%%%%
The final point concerns a further quantitative evidence of the screw-like phases. In line with previous observations, this can be also highlighted using the $g_{1w}(r_{\parallel})$ function defined in Eq.(\ref{sec1:eq9}). This is reported in Figure \ref{fig:fig9} for NPT-MC simulations with $N=2400$ helices at the same state points as in Figure \ref{fig:fig8} where the mapping between  reduced pressures and volume fractions is obtained using the equation of state Figure \ref{fig:fig4a}. Here clear, regular oscillations are visible starting at a reduced pressure $P^{*}=1.0$, roughly corresponding to $\eta=0.40$ and hence to the onset of the screw-nematic phase. The period also coincides with the pitch $p=8D$, in agreement with past studies \cite{Kolli2014a,Kolli2014b}. We also found this calculation to be less accurate when carried out at constant volume.

%%%%%%%%%%%%%%%%%%%%%%%%%%%%%%%%%%%%%%%%%%%%%%%%%%%%%%%%%%%%%%%
\subsection{Hard helices with a single attractive bead ($\chi=6.7\%$)}
\label{subsec:single}
%%%%%%%%%%%%%%%%%%%%%%%%%%%%%%%%%%%%%%%%%%%%%%%%%%%%%%%%%%%%%%

%%%%%%%%%%%%%%%%%%%%%%%%%%%% Fig 10 %%%%%%%%%%%%%%%%%%%%%%%%
\begin{figure*}[htpb]
  \centering
   \captionsetup{justification=raggedright,width=\linewidth}
     \begin{subfigure}{0.45\textwidth}
    \includegraphics[trim=0 0 0 0,clip,width=0.9\linewidth]{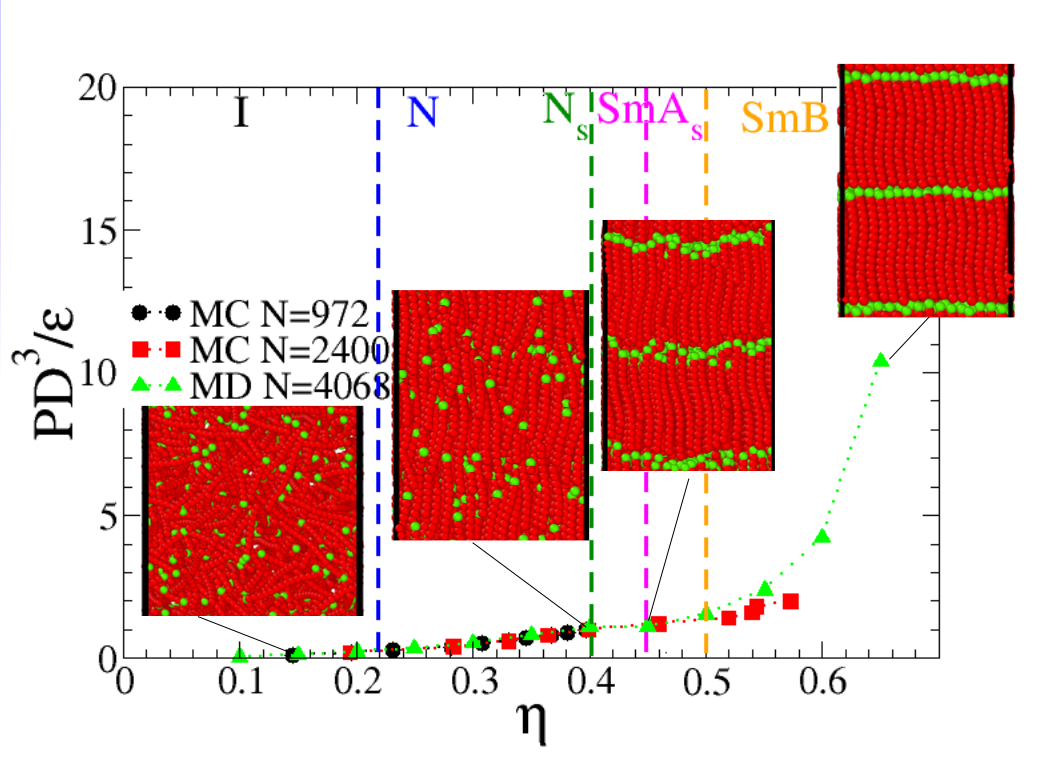}
    \caption{}\label{fig:fig10a}
   \end{subfigure}
   \begin{subfigure}{0.45\textwidth}
    \includegraphics[trim=0 0 0 0,clip,width=0.9\linewidth]{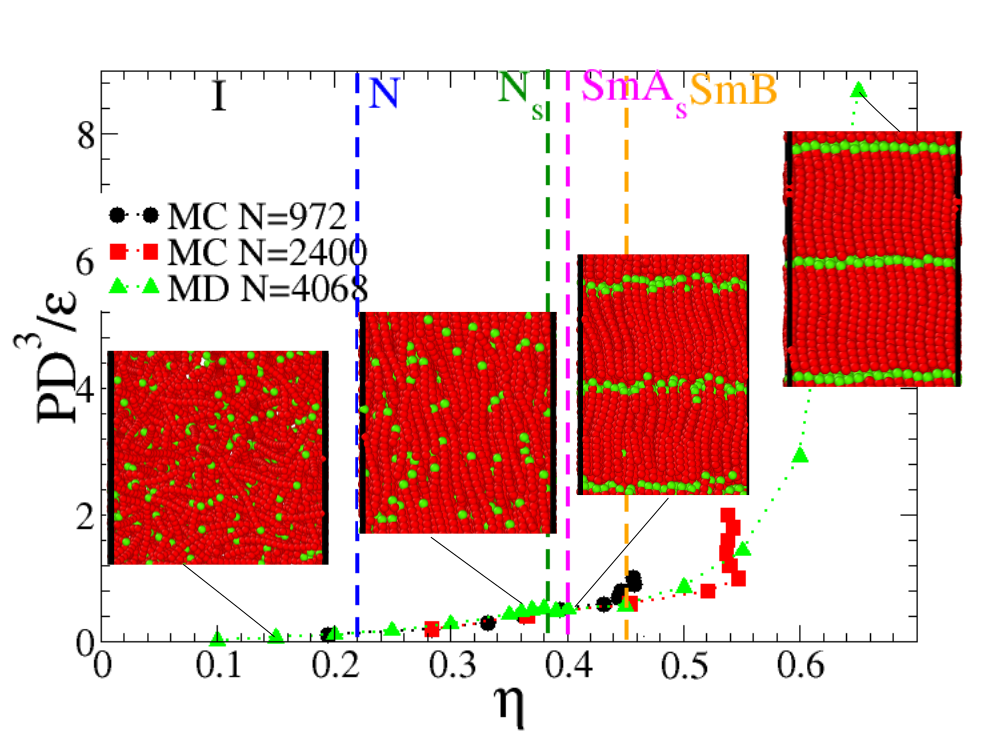}
    \caption{}\label{fig:fig10b}
   \end{subfigure} \\
   \begin{subfigure}{0.45\textwidth}
    \includegraphics[trim=0 0 0 0,clip,width=0.9\linewidth]{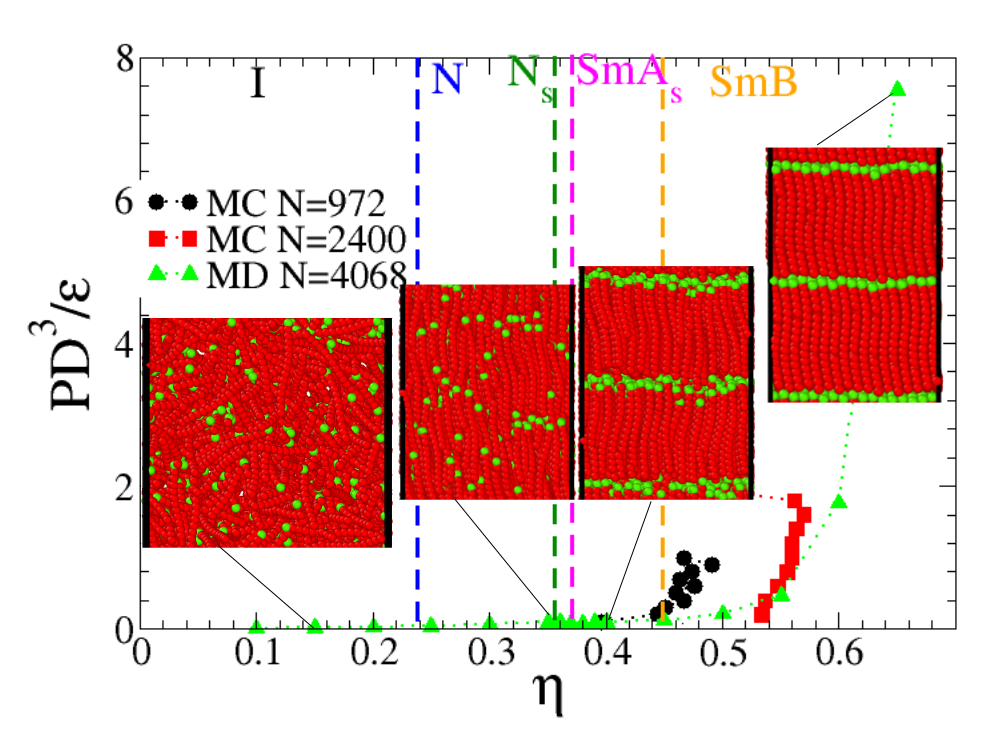}
    \caption{}\label{fig:fig10c}
   \end{subfigure}
\begin{subfigure}{0.45\textwidth}
    \includegraphics[trim=0 0 0 0,clip,width=0.9\linewidth]{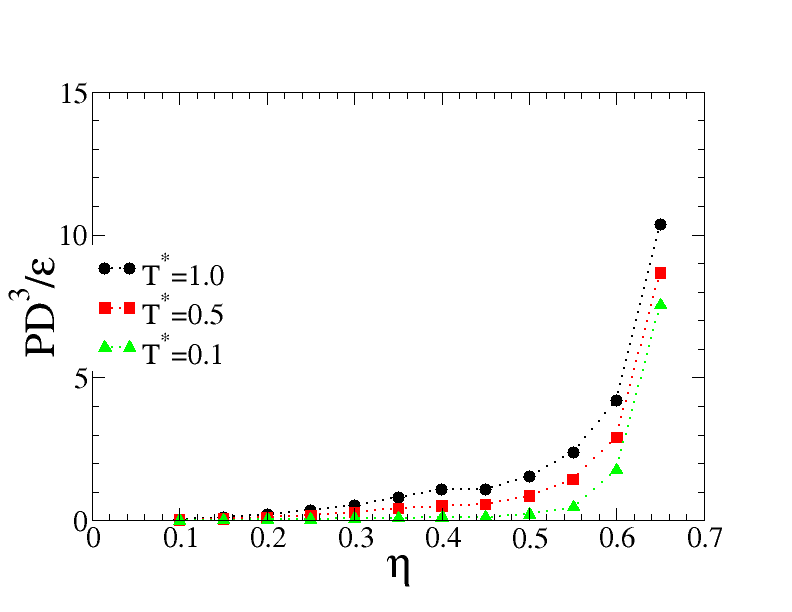}
     \caption{}\label{fig:fig10d}
   \end{subfigure}
    \caption{ Reduced pressure $P^{*}=P D^3/\epsilon$ as a function of the packing fraction $\eta$ for the case of a single attractive bead ($\chi=0.67$) at reduced temperatures $T^{*}=k_B T/\epsilon$ corresponding to (a) $T^{*}=1.0$, (b) $T^{*}=0.5$, and (c) $T^{*}=0.1$. Representative snapshots of each phase are shown for specific state points. Attractive beads are colored in green, hard (repulsive) beads in red. Results are shown for both small ($N=972$ black solid circles) and large ($N=2400$ red solid squares) NPT-MC simulations, as well as for even larger NVT-MD simulations ($N=4068$ green solid triangles). (d) Comparison of the NVT-MD results at different reduced temperatures $T^{*}=1.0,0.5,0.1$. }
   \label{fig:fig10}
 \end{figure*}
%%%%%%%%%%%%%%%%%%%%%%%%%%%%%%%%%%%%%%%%%%%%%%%%%%%%%%%%%
We now consider the case where some of the beads are attractive and apply the same machinery discussed so far for hard helices. We start by considering in detail the case of a single attractive bead, corresponding to approximately $\chi=6.7\%$ coverage (see Figure \ref{fig:fig1}).
Figure \ref{fig:fig10} reports the reduced pressure $P^{*}=P D^3 /\epsilon$ as a function of the volume fraction $\eta$ (the equation of state) for different reduced temperatures $T^{*}=k_B T/\epsilon$: (a) $T^{*}=1.0$, (b) $T^{*}=0.5$, and (c) $T^{*}=0.1$. Colored vertical bars identify the phase boundaries as was done for the hard helices case (Figure \ref{fig:fig4a}), and representative snapshots are included to highlight specific state points. Note that, at variance with the hard helices case displayed in Figure \ref{fig:fig5}, here different beads are color-coded according to their interactions, with green for attractive beads and red for purely repulsive ones. In all cases, the initial conditions have been taken as a set of parallel helices with all attractive tips aligned "up". A different choice for the initial condition will be discussed later on in Section \ref{subsec:stability}. It is worth noticing that the introduction of a fraction of attractive beads breaks the "up-down" symmetry of the original hard helices case, and introduces a preferred sense. As in the hard helices case, we report results from small ($N=972$ black solid circles) and large ($N=2400$ red solid squares) NPT-MC simulations, along with extensive NVT-MD ($N=4068$ green solid triangle up). The "small"  NPT-MC simulations are used to contrast with the original simulations with hard helices.
It is important here to stress once more that all these temperatures are well above the corresponding Boyle temperature $T_B(\chi=6.7\%) \approx 0.0022$ (see Section \ref{subsec:B2_chi}) that is nearly 2 orders of magnitude smaller compared with the minimum temperature considered here. The goal here is to study the stability of the various liquid crystal phases -- and specifically the screw phases found in the case of hard helices, against the introduction of a weak attraction.

In Figures \ref{fig:fig10a}-\ref{fig:fig10c}, we observe the same sequence of liquid crystal phases with nearly the same location of the phase boundaries as in the case of hard helices (compare with Figure \ref{fig:fig4a}). As a general rule, we observe a shift of the smectic phases to lower $\eta$ upon decreasing the temperature. In particular the onset of the \textrm{SmA}$_s$ phase is shifted from $\approx 0.45$ at $T^{*}=1.0$ (Figure \ref{fig:fig10a}) to $\approx 0.37$ at $T^{*}=0.1$ (Figure \ref{fig:fig10c}). Likewise the \textrm{SmB} phase at $T^{*}=0.1$ starts at $\approx 0.45$ (Figure \ref{fig:fig10c}) as opposed to at $\approx 0.50$ at $T^{*}=1.0$ (Figure \ref{fig:fig10a}).  The actual effect of the temperature can be inferred from Figure \ref{fig:fig10d} that summarizes the behavior of the reduced pressure $P^{*}=P D^3 /\epsilon$ as a function of the volume fraction $\eta$ for all three considered temperatures. Here, we note a decrease of the pressure at a given packing fraction $\eta$ on lowering the temperature, a trend that can be ascribed to the increasing contribution of the attractive interactions. Here the phase boundaries were located using the same methodology previously presented for fully repulsive hard helices. 

As in the fully repulsive hard helices, the \textrm{SmA}$_s$ phase is screw, in the sense that successive layers  have in-plane positional disorder of the center-of-mass of the helices in that plane, but aligned helix secondary axes $\widehat{\mathbf{w}}$ along the phase secondary axis $\widehat{\mathbf{C}}$ that rotate for successive layers along $\widehat{\mathbf{N}}$ (see Figure \ref{fig:fig2b}). Hence, the nematic region is split in two sub-regions, a conventional nematic \textrm{N} -- presumably a cholesteric one when seen at a larger scale, at lower volume fraction, and a screw-nematic \textrm{N}$_{s}$ at higher volume fraction whose boundary is only mildly dependent on the temperature.

Supplementary Movie2 shows evidence of the screw-nematic phase for the single attractive bead along the lines used in the case of fully repulsive hard helices (Movie1), and Supplementary Figure SVI provides the corresponding relevant snapshots, the counterpart of Figure \ref{fig:fig7}.

Also worth noting is the increasing tendency of the NPT-MC simulations to become kinetically trapped. This is especially visible at $T^{*}=0.1$ (Figure \ref{fig:fig10c}), where the pressure in both the $N=972$ and $N=2400$ NPT-MC simulations deviates substantially from the NVT-MD results at high volume fraction. This deviation is likely related to the difficulty of the NPT-MC runs in reaching full equilibration and thermalization (see Supplementary Figure SIII). This issue becomes even more severe as the fraction of attractive beads increases, and hence NPT-MC results will not be discussed further in this study.

This single attractive bead case is particularly interesting because at very low temperatures (lower that the Boyle temperature $T_B(\chi=6.7\%) \approx 0.0022$), we expect the formation of micelles at low densities replacing the isotropic phase. This is indeed what occurs for Janus rods with a single attractive site (see companion paper \cite{Wood2023}). In contrast, at the temperatures considered in the present study, the liquid crystal phases are modified but not destabilized as the attractive energy is insufficient to compensate for the higher entropy of the thermotropic liquid crystal phases. Hence, the phase behavior of these systems is still mainly determined by entropy. At lower temperature, however, there should be a transition region in which these two opposite tendencies compete and give rise to interesting effects. This can in fact be seen in the companion paper on Janus rods \cite{Wood2023}. 
%%%%%%%%%%%%%%%%%%%%%%%%%%%%%%%%%%%%%%%%%%%%%%%%%%%%%%%%%%%%%%
\subsection{Stability with respect to the initial conditions}
\label{subsec:stability}
%%%%%%%%%%%%%%%%%%%%%%%%%%%%%%%%%%%%%%%%%%%%%%%%%%%%%%%%%%%%%%
%%%%%%%%%%%%%%%%%%%%% FIG 11 %%%%%%%%%%%%%%%%%%%%%%%%%%%%%%%%%
\begin{figure}
    \centering
    \includegraphics[trim=0 0 0 0,clip,width=1.0\linewidth]{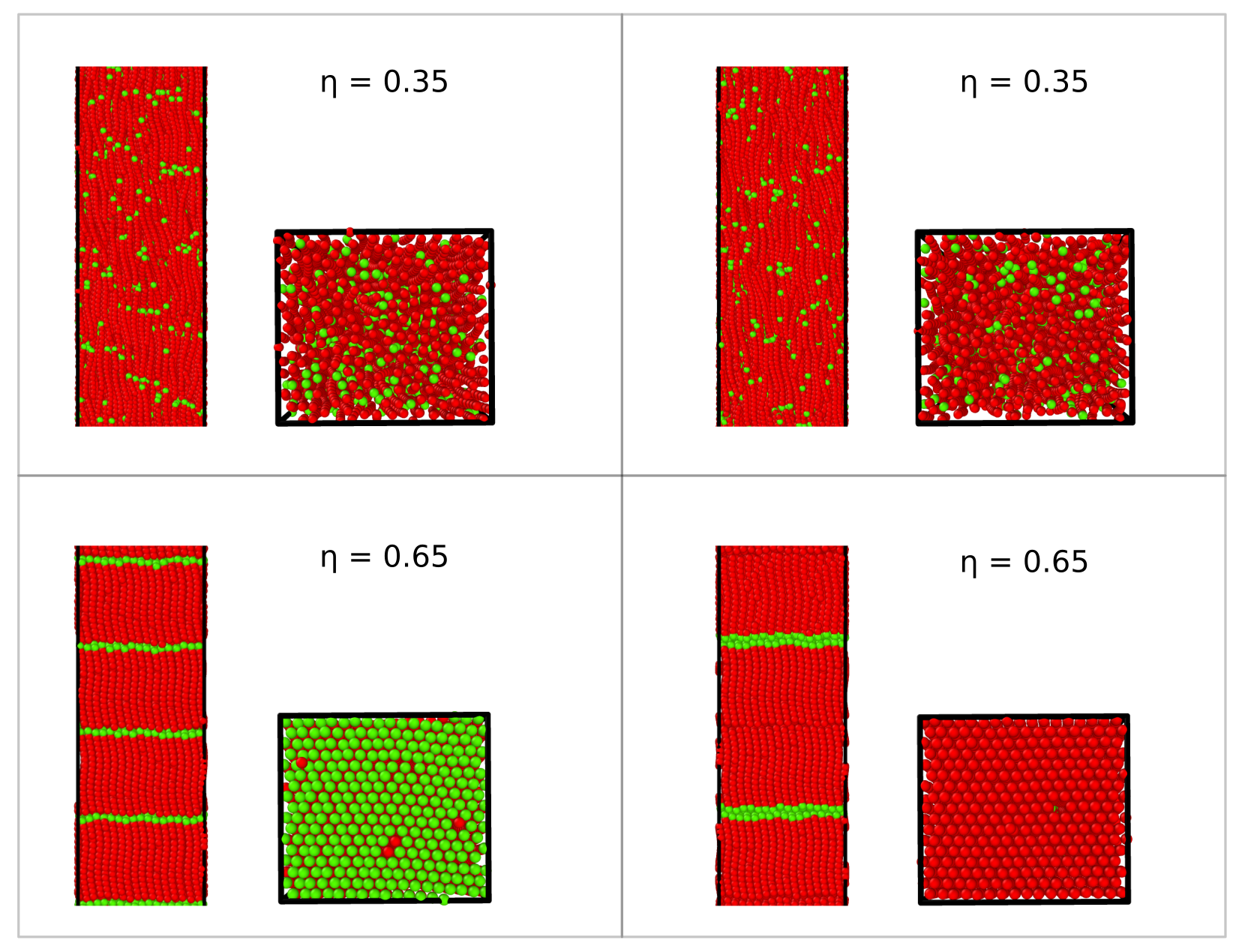}
    \caption{Test of the stability with respect to the initial conditions. Top panel left: final equilibrated conformation at $\eta=0.35$ (nematic phase) obtained with all helices initially parallel (i.e. with attractive tips all pointing along the same direction); Top panel right: Same but with antiparallel initial conditions (attractive tips up for odd layers and down for even layers); Bottom panel left: Same as before with $\eta=0.65$ (SmB phase); Bottom panel right: Same with antiparallel initial conditions. In all cases, the lowest temperature $T^{*}=0.1$ has been considered.}
    \label{fig:fig11}
\end{figure}
%%%%%%%%%%%%%%%%%%%%%%%%%%%%%%%%%%%%%%%%%%%%%%%%%%%%%%%%%%%%%%%%%
In principle, it would be desirable to have a final configuration that is fully independent of the initial conditions. In practice, however, this is hardly achievable for the most compact configurations, even in the absence of attractive interactions \cite{Kolli2014b} or for hard spherocylinders \cite{Bolhuis1997}. We therefore explicitly checked for this dependence and report it in Figure \ref{fig:fig11} by considering two different initial conditions. In the first case, used in all of the cases discussed so far, all helices are oriented with the attractive beads initially pointing up (i.e. along the $+z$ direction). We denote this as the \textit{parallel} initial condition. In the second case, the odd layers have initially helices oriented with the attractive beads pointing up, whereas in the even layers all helices have their attractive beads initially pointing down (along the $-z$ direction). We denote this as \textit{antiparallel} initial condition.

In the top panels of Figure \ref{fig:fig11}, we show two equilibrated configurations obtained at the same volume fraction of $\eta =0.35$ and at the same temperature $T^{*}=0.1$ (corresponding to a nematic \textrm{N} phase), but starting with parallel (left) and antiparallel (right) initial conditions. Very reassuring, the two final configurations are essentially indistinguishable and hence equivalent from a statistical viewpoint. However, this turns out not to be the case for the most demanding case of $\eta=0.65$ (bottom panel) corresponding to a SmB phase, where it is clear that initial parallel (left) and antiparallel (right) configurations are essentially preserved upon equilibration to the local thermodynamically stable states. As mentioned, this is a very common feature of particles with a significant aspect ratio, and it should not come as a surprise, since we have also considered the most challenging situation of $T^{*}=0.1$. We further note that the total energy of the antiparallel conformation (right) is lower than the parallel one (left) at temperature $T^{*}=0.1$, thus indicating that the antiparallel configuration becomes more stable than the parallel one as the temperature decreases. In our companion paper \cite{Wood2023} on Janus rods, the antiparallel layered configuration appears spontaneously upon compression. These findings suggest that the screw-nematic phases that we discussed so far should not be affected by the energetic preference of the antiparallel conformation, but the smectic ones might.
%%%%%%%%%%%%%%%%%%%%%%%%%%%%%%%%%%%%%%%%%%%%%%%%%%%%%%%%%%%%%%%
\subsection{Janus ($\chi=50\%$) and fully attractive ($\chi=100\%$) helices}
\label{subsec:janus}
%%%%%%%%%%%%%%%%%%%%%%%%%%%%%%%%%%%%%%%%%%%%%%%%%%%%%%%%%%%%%%
%%%%%%%%%%%%%%%%%%%%%%%%%%%%%% FIG 12 %%%%%%%%%%%%%%%%%%%%%%%%%%%%%%%%%%%%%%%%%%%
\begin{figure*}
   \begin{subfigure}[b]{8cm}
    \includegraphics[trim= 10 10 10 10,clip,width=0.9\linewidth]{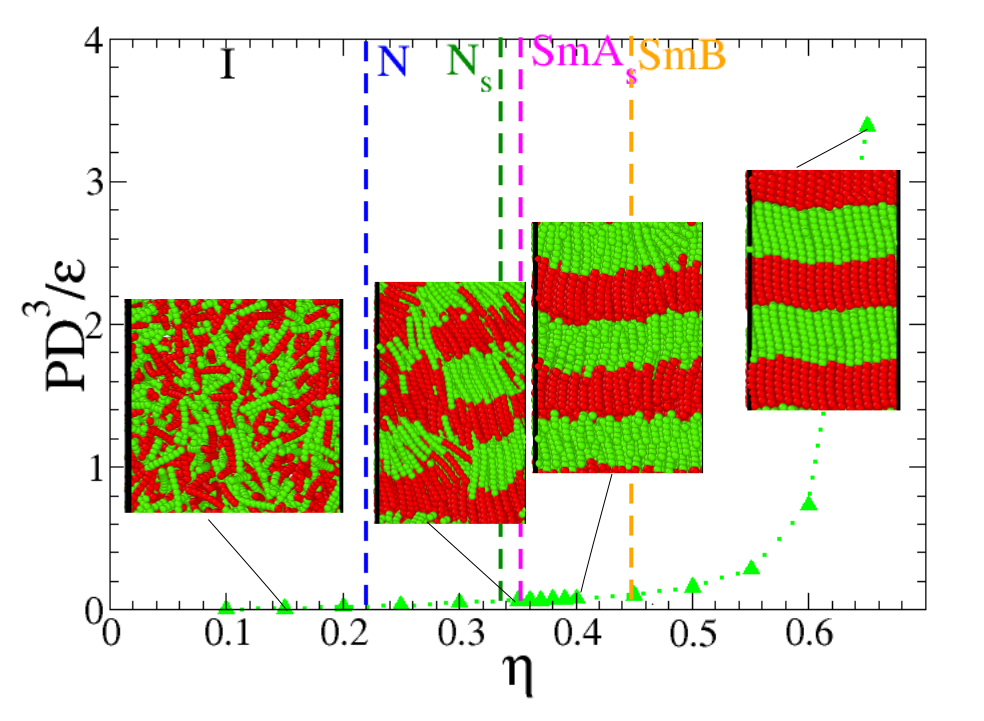}
    \caption{}\label{fig:fig12a}
    \end{subfigure}
    \hfill
    \begin{subfigure}[b]{8cm}
    \includegraphics[trim= 0 0 0 0,clip,width=1.0\linewidth]{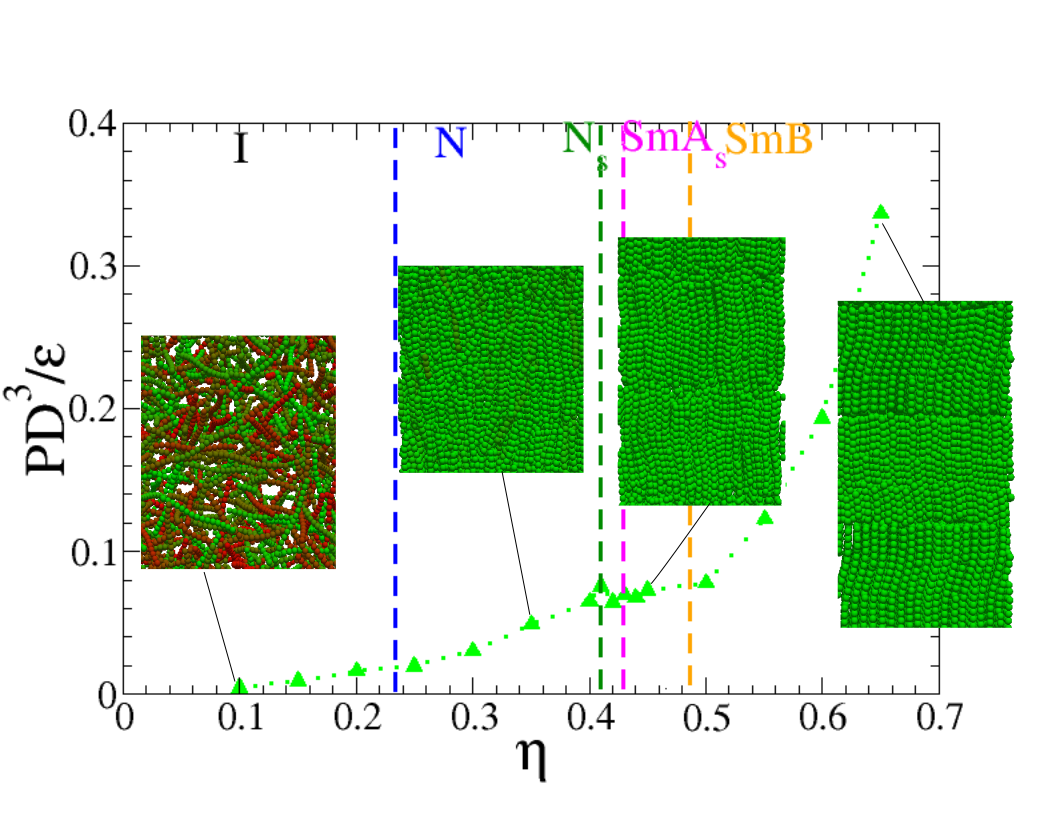}
    \caption{ }\label{fig:fig12b}
   \end{subfigure}
   \hfill
  \caption{Reduced pressure $P D^3/\epsilon$ as a function of packing fraction $\eta$ from MD-NVT simulations with $N=4068$ helices at $T^{*}=0.1$. (a) Case of Janus helices (8/15 attractive beads, $\chi=50\%$). Green = attractive beads, Red = hard core beads. (b) Case of fully attractive (SW) helices ($\chi=100\%$). Here helices are colored according to their orientations. Representative snapshots at indicated state points are also displayed. }
  \label{fig:fig12}
\end{figure*}
%%%%%%%%%%%%%%%%%%%%%%%%%%%%%%%%%%%%%%%%%%%%%%%%%%%%%%%%%
It is instructive to inspect what happens to the phase behavior upon increasing the number of attractive beads (i.e. the coverage $\chi$). Two cases appear to be particularly interesting. 

The first case is when approximately half of the beads of the helices are attractive (8 out of 15), which we refer to as the Janus limit ($\chi \approx 50\%$, see Figure \ref{fig:fig1}). The $\chi=50\%$ limit has shown a particularly rich phenomenology both, in the case of single colloids \cite{Sciortino2009,Sciortino2010} and in the case of dumbbells \cite{OToole2017a,OToole2017b}. 
At very low temperatures -- lower than the corresponding Boyle temperature 
$T_B^{*} (\chi \approx 50\%)\approx 0.015$ --
one might expect the formation of bilayers where two attractive halves of two helices bind to form a non-covalent bond. This is indeed what happens for Janus rods (see the companion paper \cite{Wood2023}). However, as in the single attractive bead case ($\chi=6.7\%$) discussed earlier in Section \ref{subsec:single}, the range of temperatures considered in this work, $0.1 \le T^{*} \le 1.0$,
is too high to destabilize the entropically dominated phases found in the hard helices counterpart. This is clearly visible in Figure \ref{fig:fig12a}), which reports the equation of state at $T^{*}=0.1$. Compared to the single attractive bead counterpart (see Figure \ref{fig:fig10c}), the phase boundaries are all at very similar positions. Supplementary Figure SIX shows the potential energy per helix $E_p/N$ as a function of the volume fraction $\eta$ from the NVT-MD simulations at $T^{*}=0.1$. As expected, for nearly all coverages $\chi$ this potential energy is positive and increases with increasing $\eta$. Not surprisingly, in the case of fully attractive helices $\chi=100\%$ case, the energy is slightly negative and decreases on increasing $\eta$.

The second case is the extension of the attractive well to all beads (SW $\chi = 100\%$). This leads to the equation of state displayed in Figure \ref{fig:fig12b}, again for $T^{*}=0.1$. Unlike the partially attractive cases of $\chi=6.7\%$ and $\chi \approx 50\%$, but in line with purely hard helices, the inserted snapshots representing selected state points are here colored according to the direction $\widehat{\mathbf{u}}$ of the main axis compared with the main director $\widehat{\mathbf{N}}$. This case is particularly interesting because of its dual nature. On the one hand, it represents a smooth extension of the previous $\chi=6.7\%$ and $\chi \approx 50\%$ cases in terms of coverage. However, in contrast to the $\chi=6.7\%$ and $\chi \approx 50\%$ cases, each helix can be flipped upside-down with no effect on the mesophase, thus restoring the up-down symmetry. A comparison with the Janus case (Fig. \ref{fig:fig12a}) shows that all transitions have shifted to slightly higher volume fraction, increasing their region of stability and nearly reproducing the same results of the fully repulsive hard helices (see Figure \ref{fig:fig4a}).  This "re-entrant" behaviour can likely be ascribed to restoring the up-down symmetry as remarked.

%%%%%%%%%%%%%%%%%%%%%%%%%%%%%%%%%%%%%%%%%%%%%%%%%%%%%%%%%%%%%%
\subsection{Temperature and coverage dependence}
\label{subsec:temperature}
%%%%%%%%%%%%%%%%%%%%%%%%%%%%%%%%%%%%%%%%%%%%%%%%%%%%%%%%%%%%%%

%%%%%%%%%%%%%%%%%%%%%%% FIG 13 %%%%%%%%%%%%%%%%%%%%%%%%%%%%%%%%%
\begin{figure}
    \centering
    \includegraphics[trim=10 10 10 10,clip,width=1.0\linewidth]{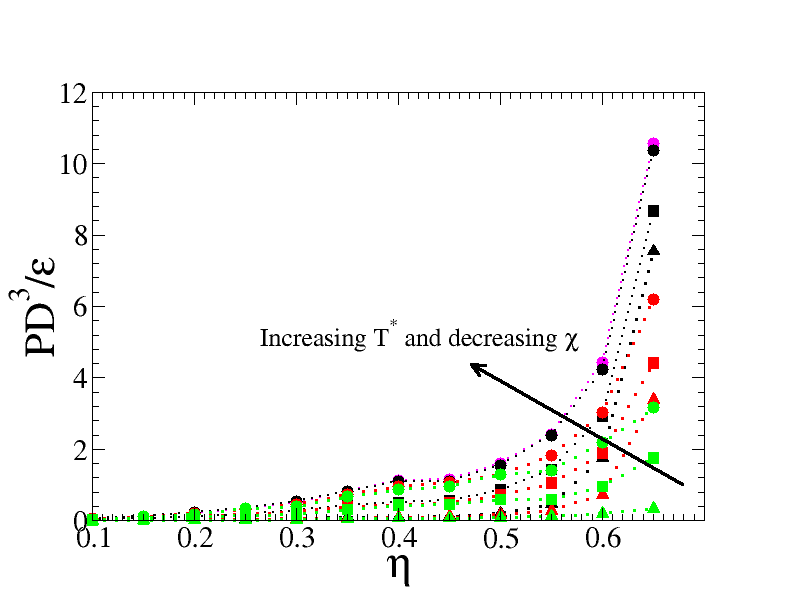}
    \caption{Reduced pressure $P^{*}=P D^3/\epsilon$ as a function of the packing fraction from MD simulations at different reduced temperatures $T^{*}=k_B T/\epsilon=1.0$ (solid circles), $T^{*}=k_B T/\epsilon=0.5$ (solid squares), and $T^{*}=k_B T/\epsilon=0.1$ (solid triangles). Different  coverages  $\chi=6.7 \%$ single attractive bead (black), $\chi=50\%$ Janus case (red), and $\chi=100\%$ SW case (green) are also shown. For reference, the hard helices case $\chi=0\%$ (solid circle magenta) is also reported. In this case, the natural unit $P^{*} = P D^3 /k_B T$ has been used for the pressure. }
    \label{fig:fig13}
\end{figure}
%%%%%%%%%%%%%%%%%%%%%%%%%%%%%%%%%%%%%%%%%%%%%%%%%%%%%%%%%%%%%%%%%
We can now summarize our findings on the equation of state, and the location of the transitions, in terms of the temperature and coverage dependence. Figure \ref{fig:fig13} reports such a comparison for coverages from hard to fully attractive and temperatures from $T^{*}=1.0$ to $T^{*}=0.1$. This shows that the pressure increases upon increasing the temperature (at a given volume fraction and coverage) or upon decreasing the coverage $\chi$ (at a fixed temperature). Again, this can be rationalized on the basis that decreasing the temperature and/or increasing the fraction of attractive beads increase self-attraction and hence decreases the pressure. How the phase boundaries depend on coverage is more complex. At the lowest considered temperature $T^{*}=0.1$, the phase boundaries of the partial coverage cases (single attractive bead and Janus helices) do not display any dependence on the coverage. Somewhat surprisingly, the phase boundaries of the fully attractive case appear to have nearly identical locations as the case of hard helices.
A summary of the coverage dependence of the various transitions can be found in Supplementary Figure SX.

A progressive destabilization of the isotropic \textrm{I} phase upon cooling was previously observed in a system of square-well prolate spherocylinders with a continuous line of interaction sites \cite{Gamez2017}. Note however that in that case the \textrm{I} boundary does shift, but not because the nematic \textrm{N} and smectic \textrm{Sm} phases become more stable. Their phase diagram shows that the \textrm{Sm} phase disappears at low $T$ and that the shift in the \textrm{I} boundary is due to growth in size of a coexistence region rather than the \textrm{N} phase. We further note that study cannot be compared quantitatively with the present one as $L/D=5$ in that case. This point is further discussed in the companion paper \cite{Wood2023}, which considers Janus rods with $L/D=5$.
%%%%%%%%%%%%%%%%%%%%%%%%%%%%%%%%%%%%%%%%%%%%%%%%%%%%%
\section{Conclusions}
\label{sec:conclusions}
%%%%%%%%%%%%%%%%%%%%%%%%%%%%%%%%%%%%%%%%%%%%%%%%%%%%%
In this study we considered the phase behavior of a system of helices formed by a set of fused beads arranged into a prescribed helical shape. Unlike previous studies that focused on hard helices \cite{Kolli2014a,Kolli2014b,Kolli2016,Cinacchi2017}, here some of the beads on different helices attract one another, with  the fraction of attractive sites ranging from $100\%$ coverage to $0\%$ (pure hard helices). Using this system, we attempted to address the following two questions: (1) How does attraction change the phase behavior from that of hard helices?; and (2) What regions of the parameter space are worth studying in more depth? 

Starting from the known phase diagram of the hard helices case, we studied the effects of coverage and temperature on the phase behavior. This was done in regimes well above the Boyle temperature but at sufficiently low temperature (in some cases) for attraction to compete with the purely entropic effects that determine the ordering of hard helices. 
As the Boyle temperature for low coverage (e.g. $T_B^{*} \approx 0.0015$ in the case of a single attractive site) is considerably smaller than at high coverage (e.g. $T_B^{*} \approx 0.04$ for a system of fully attractive hard helices), the results obtained at the considered temperatures ($T^{*}=0.1,0.5,1.0$) probe different multiples of $T_B^{*}$, but with the common denominator of being all in the entropically dominated regime.

In this work, we focused on a slender helical shape with radius $R=0.2D$ and pitch $p=8D$ on the basis that this is nearly a rod-like particle but still chiral. We also extended past studies of hard helices in terms of the number of particles, the maximum volume fraction, and by including a study of the particle dynamics via the use of molecular dynamics simulations. This allowed us to check previous results and revealed possible kinetic trapping in MC simulations stemming from the combined effects of hard helical interactions, high aspect ratio, low temperature and high packing fraction. We found that while fully consistent for hard helices and high temperatures, NPT-MC calculations become problematic at low temperatures and very high densities because of slow equilibration and kinetic trapping. For this reason, in the present study we mainly used molecular dynamics simulations with slightly softer potentials, which allow for parallel simulation, in addition to providing access to particle dynamics.

In the case of hard helices, we confirmed previous results \cite{Frezza2013,Frezza2014,Kolli2014a,Kolli2014b,Kolli2014b,Cinacchi2017} and provided further evidence on the presence of screw-like phases, both nematic and smectic, originating from the helical shape. We then extended the model to include attraction between a fraction of the beads forming each helix, and studied the stability of the various phases upon lowering the temperature and increasing the fraction of attractive beads. In all cases, the range of temperatures selected were well above the corresponding Boyle temperatures, meaning that the attractive interactions add an energetic component to the free energy but that the phase behaviour is still dominated by entropy. As a result, we found a pressure decrease at fixed volume fraction on lowering the temperature and/or increasing the fraction of attractive beads, a fact that can be easily rationalized with the progressive increase of the relative balance between attraction and repulsion, but only small changes in the phase behavior. While the location of the nematic phases are only mildly affected, we observed a shift of the screw and the smectic phases to lower volume fraction. However, this shift appears to be non-monotonic, with the fully attractive helices behaving differently from the partially attractive counterparts. For the case of a single attractive bead, we observed no tendency of the liquid crystal phases to compete with the formation of micelles at the considered temperatures. We do expect the onset of micelles  below the Boyle temperature at sufficiently low densities, as this is what occurs for Janus rods studied in the companion paper \cite{Wood2023}. Using the same rationale, we expect Janus helices where half of the beads are attractive to display competition between the formation of lamellar phases and the liquid crystal phases observed in this work at sufficiently low temperature, again discussed in the companion paper on Janus rods \cite{Wood2023}.

There are many avenues that the present study (as well as the companion paper on Janus rods \cite{Wood2023}) open for future analyses. As this study focused on very slender helices, it would be interesting to check what happens to much curlier helices where the in-plane locking of neighboring parallel helices is expected to be much more effective at reducing the rotational entropy, thus further promoting the screw-like mechanism which stabilizes the screw-like phases observed here. Another point which deserves further attention is the effect of attraction on the cholesteric phase observed for hard helices \cite{Frezza2014,Cinacchi2017}. Finally, a recent study by some of the current authors \cite{Liu2022} has highlighted the important role that helical shape and chirality has on the twisting of monolayer assemblies of rod-like or helical particles. It would therefore be interesting to extend the low temperature analysis carried out for Janus rods to Janus helices, thus probing the complementary regime for helices. We plan to pursue these investigations in a future study.

\section*{Supplementary Material}
See the supplementary material for additional results in the case of fully repulsive hard helices and hard helices with a single attractive beads and for representative movies highlighting the screw-like nematic phase in these two cases.

\begin{acknowledgments}
The present study builds on a past collaboration that was initiated together with Alberta Ferrarini and Giorgio Cinacchi.
The use of the SCSCF multiprocessor cluster at the Universit\`{a} Ca' Foscari Venezia is gratefully acknowledged. The authors acknowledge financial support by  MIUR PRIN-COFIN2022 grant 2022JWAF7Y (AG), the Australian Research Council Grants CE170100026 and FT140101061 (AWC and JW), the Galileo Project 2018-39566PG (AG), and the Erasmus+ International Mobility Program. 
\end{acknowledgments}

\section*{Data Availability Statement}
The data that support the findings of this study are available within the article [and its Supplementary Material].

\section*{Author declaration}
The authors have no conflict to disclosure

\section*{Author contributions}
LDC: software, formal analysis, investigation, methodology, review $\&$ editing. FR: methodology, review $\&$ editing. JAW: software, formal analysis, methodology. AWC: conceptualization, funding acquisition, review $\&$ editing. AG: conceptualization, software, formal analysis, methodology, funding acquisition, review $\&$ editing.

%\nocite{*}
\bibliography{jcp_arxiv}% Produces the bibliography via BibTeX.

\end{document}